\documentclass[preprint,aps,prd,floatfix,nofootinbib,superscriptaddress,eqsecnum,tightenlines]{revtex4-1}

\usepackage[mathscr]{euscript}
\usepackage{amsmath,amsfonts,amssymb}
\usepackage{bm}
\usepackage{graphicx}
\usepackage[pdftex]{color}
\usepackage[sort&compress]{natbib}
\usepackage[colorlinks=true,linkcolor=blue,filecolor=blue,urlcolor=blue,citecolor=blue,pdftex,plainpages=false]{hyperref}

\newcommand{\eq}[1]{Eq.~\eqref{eq:#1}}

\newcommand{\eqs}[2]{Eqs.~\eqref{eq:#1} and \eqref{eq:#2}}

\def\eqn#1{\label{eq:#1}}

\def\figref#1{Fig.~\ref{fig:#1}}
\def\Figref#1{Figure~\ref{fig:#1}}
\def\figrefs#1#2{Figs.~\ref{fig:#1} and~\ref{fig:#2}}
\def\Figrefs#1#2{Figures~\ref{fig:#1} and~\ref{fig:#2}}
\def\secref#1{Sec.~\ref{sec:#1}}

\def\Secref#1{Section~\ref{sec:#1}}
\def\tabref#1{Table~\ref{tab:#1}}
\def\rcite#1{Ref.~\cite{#1}}

\def\rcites#1{Refs.~\cite{#1}}

\newcommand{\ie}{\emph{i.e.},\ }
\newcommand{\eg}{\emph{e.g.},\ }

\newcommand{\chpt}{$\chi$PT}
\newcommand{\hmchpt}{HM$\chi$PT}
\newcommand{\schpt}{S$\chi$PT}

\newcommand{\aschpt}{HMrAS$\chi$PT}

\newcommand{\cC}{\ensuremath{\mathcal{C}}}

\newcommand{\cF}{\ensuremath{\mathcal{F}}}

\newcommand{\cJ}{\ensuremath{\mathcal{J}}}

\newcommand{\cM}{\ensuremath{\mathcal{M}}}

\newcommand{\order}{\ensuremath{\text{O}}} 

\newcommand{\param}[1]{{\textcolor{blue}{#1}}} 

\newcommand{\half}{\ensuremath{{\textstyle\frac{1}{2}}}}
\newcommand{\third}{\ensuremath{{\textstyle\frac{1}{3}}}}
\newcommand{\quarter}{\ensuremath{{\textstyle\frac{1}{4}}}}

\newcommand{\BE}{\begin{displaymath}}
\newcommand{\EE}{\end{displaymath}}
\newcommand{\BNE}{\begin{equation}}
\newcommand{\ENE}{\end{equation}}
\newcommand{\BEA}{\begin{eqnarray}}
\newcommand{\EEA}{\nonumber\end{eqnarray}}

\newcommand{\leftvec}{{\raise1.5ex\hbox{$\leftarrow$}\kern-1.00em}}
\newcommand{\rightvec}{{\raise1.5ex\hbox{$\rightarrow$}\kern-1.00em}}

\newcommand{\GeV}{\text{GeV}}
\newcommand{\MeV}{\text{MeV}} 
\newcommand{\fpiPDG}{\ensuremath{f_{\pi,\text{PDG}}}}

\newcommand{\LamQCD}{\ensuremath{\Lambda_\text{QCD}}}
\newcommand{\LamHQET}{\ensuremath{\Lambda_\text{HQET}}}
\newcommand{\pole}{{\text{pole}}}
\newcommand{\RS}{{\text{RS}}}
\newcommand{\MRS}{{\text{MRS}}}
\newcommand{\MSbar}{{\ensuremath{\overline{\text{MS}}}}}
\newcommand{\Lambdabar}{{\ensuremath{\overline{\Lambda}}}}
\newcommand{\mbar}{\overline{m}}
\newcommand{\mcbar}{{\overline{m}_c}}
\newcommand{\mhbar}{{\overline{m}_h}}
\newcommand{\mbbar}{{\overline{m}_b}}

\newcommand{\vt}{x} 

\def\rHISQrCONFIGS{Bazavov:2012xda}
\def\rFD2014{Bazavov:2014wgs, *Bazavov:2014lja}
\def\rPDG2016{Olive:2016xmw, *Rosner:2015wva}

\newcommand{\EMone}{$K^+$-$K^0$ splitting}
\newcommand{\EMtwo}{$K^0$~mass}

\newcommand{\EMthree}{$H_x$~mass}

\allowdisplaybreaks

\begin{document}

\title{Up-, down-, strange-, charm-, and bottom-quark masses from four-flavor lattice QCD}

\author{A.~Bazavov} 
\affiliation{Department of Computational Mathematics, Science and Engineering,
\\ and Department of Physics and Astronomy, Michigan State University, East Lansing, Michigan 48824, USA}

\author{C.~Bernard}
\email[]{cb@lump.wustl.edu}
\affiliation{Department of Physics, Washington University, St. Louis, Missouri 63130, USA}

\author{N.~Brambilla}
\email[]{nora.brambilla@ph.tum.de}
\affiliation{Physik-Department, Technische Universit\"at M\"unchen, 85748 Garching, Germany}
\affiliation{Institute for Advanced Study, Technische Universit\"at M\"unchen, 85748 Garching, Germany}

\author{N.~Brown}
\affiliation{Department of Physics, Washington University, St. Louis, Missouri 63130, USA}

\author{C.~DeTar}
\affiliation{Department of Physics and Astronomy, University of Utah, Salt Lake City, Utah 84112, USA}

\author{A.X.~El-Khadra}
\affiliation{Department of Physics, University of Illinois, Urbana,  Illinois 61801, USA}
\affiliation{Fermi National Accelerator Laboratory, Batavia, Illinois 60510, USA}

\author{E.~G\'amiz} 
\affiliation{CAFPE and Departamento de F\'isica Te\'orica y del Cosmos, Universidad de Granada, E-18071 Granada, Spain}

\author{Steven~Gottlieb}
\affiliation{Department of Physics, Indiana University, Bloomington, Indiana 47405, USA}

\author{U.M.~Heller} 
\affiliation{American Physical Society, One Research Road, Ridge, New~York 11961, USA}

\author{J.~Komijani}
\email[]{javad.komijani@glasgow.ac.uk}
\affiliation{Physik-Department, Technische Universit\"at M\"unchen, 85748 Garching, Germany}
\affiliation{Institute for Advanced Study, Technische Universit\"at M\"unchen, 85748 Garching, Germany}
\affiliation{School of Physics and Astronomy, University of Glasgow, Glasgow G12 8QQ, United~Kingdom}

\author{A.S.~Kronfeld} 
\email[]{ask@fnal.gov}
\affiliation{Fermi National Accelerator Laboratory, Batavia, Illinois 60510, USA}
\affiliation{Institute for Advanced Study, Technische Universit\"at M\"unchen, 85748 Garching, Germany}

\author{J.~Laiho}  
\affiliation{Department of Physics, Syracuse University, Syracuse, New~York 13244, USA}

\author{P.B.~Mackenzie}
\affiliation{Fermi National Accelerator Laboratory, Batavia, Illinois 60510, USA}

\author{E.T.~Neil}
\affiliation{Department of Physics, University of Colorado, Boulder, Colorado 80309, USA}
\affiliation{RIKEN-BNL Research Center, Brookhaven National Laboratory, \\ Upton, New York 11973, USA}

\author{J.N.~Simone}
\affiliation{Fermi National Accelerator Laboratory, Batavia, Illinois 60510, USA}

\author{R.L.~Sugar}
\affiliation{Department of Physics, University of California, Santa Barbara, California 93106, USA}

\author{D.~Toussaint}
\email[]{doug@physics.arizona.edu}
\affiliation{Physics Department, University of Arizona, Tucson, Arizona 85721, USA}

\author{A.~Vairo}
\email[]{antonio.vairo@tum.de}
\affiliation{Physik-Department, Technische Universit\"at M\"unchen, 85748 Garching, Germany}

\author{R.S.~Van~de~Water}
\affiliation{Fermi National Accelerator Laboratory, Batavia, Illinois 60510, USA}

\collaboration{Fermilab Lattice, MILC, and TUMQCD Collaborations}
\noaffiliation

\date{\today}

\begin{abstract} 
We calculate the up-, down-, strange-, charm-, and bottom-quark masses using the MILC highly improved staggered-quark ensembles with
four flavors of dynamical quarks.
We use ensembles at six lattice spacings ranging from $a\approx0.15$~fm to $0.03$~fm and with both physical and unphysical values of
the two light and the strange sea-quark masses.
We use a new method based on heavy-quark effective theory (HQET) to extract quark masses from heavy-light pseudoscalar meson masses.
Combining our analysis with our separate determination of ratios of light-quark masses we present masses of the up, down, strange,
charm, and bottom quarks.
Our results for the $\overline{\text{MS}}$-renormalized masses are %
    $m_u(2~\text{GeV}) = 2.130(41)$~MeV,
    $m_d(2~\text{GeV}) = 4.675(56)$~MeV,
    $m_s(2~\text{GeV}) = 92.47(69)$~MeV,
    $m_c(3~\text{GeV}) = 983.7(5.6)$~MeV,
and
    $m_c(m_c) = 1273(10)$~MeV,
with four active flavors;
and
    $m_b(m_b) = 4195(14)$~MeV with five active flavors.
We also obtain ratios of quark masses
    $m_c/m_s = 11.783(25)$,
    $m_b/m_s = 53.94(12)$, and
    $m_b/m_c = 4.578(8)$. %
The result for $m_c$ matches the precision of the most precise calculation to date, and the other masses and all quoted ratios are
the most precise to date.
Moreover, these results are the first with a perturbative accuracy of $\alpha_s^4$.
As byproducts of our method, we obtain the matrix elements of HQET operators with dimension 4 and 5:
$\overline{\Lambda}_\text{MRS}=555(31)$~MeV in the minimal renormalon-subtracted (MRS) scheme,
$\mu_\pi^2 = 0.05(22)~\text{GeV}^2$, and
$\mu_G^2(m_b)=0.38(2)~\text{GeV}^2$.
The MRS scheme [Phys. Rev. \textbf{D97}, 034503 (2018), arXiv:1712.04983 [hep-ph]] is the key new aspect of our method.
\end{abstract}

\preprint{FERMILAB-PUB-17/492-T}
\preprint{TUM-EFT 107/18}

\maketitle

\section{Introduction}
\label{sec:Introduction}

Quark masses are fundamental parameters of QCD.
They must be known accurately for precise theoretical calculations within the Standard Model, especially for testing whether quarks
receive mass via Yukawa couplings to the Higgs field.
Because of confinement, the quark masses can be defined only as renormalized parameters of the QCD Lagrangian.
Thus, they must be determined by comparing theoretical calculations of an appropriate set of observables to experimental
measurements of those observables.
Lattice QCD makes it possible to calculate in a nonperturbative way simple observables, such as hadron masses.
To determine the quark masses in lattice QCD, one needs to tune the bare lattice quark masses such that a suitable set of hadron
masses coincide with their experimental values.

The resulting bare masses must be renormalized, preferably to a regularization independent scheme, such as the recently introduced
minimal renormalon subtracted (MRS) mass~\cite{Brambilla:2017mrs}.
One approach is to use lattice perturbation theory, but multiloop calculations are difficult so, in practice, nothing more than
two-loop matching~\cite{Mason:2005bj,Skouroupathis:2008mf,Constantinou:2016ieh} is available in the literature.
Another is to use nonperturbative renormalization to, for example, momentum-subtraction~\cite{Martinelli:1994ty,Lytle:2013qoa} or
finite-volume~\cite{Capitani:1998mq,DellaMorte:2005kg} schemes.
Finally, one can use lattice gauge theory to obtain quantities in continuum QCD and apply multiloop continuum perturbative QCD to
extract the quark masses.
An example of the latter is the analysis of quarkonium correlators~\cite{Allison:2008xk}.
In practice, no regularization-independent scheme is in such common use as the modified minimal subtraction (\MSbar)
scheme~\cite{Bardeen:1978yd} of dimensional regularization, so we shall use \MSbar\ to quote results.
%
\phantom{\cite{Freeland:2006nd,*Freeland:2007wk}}
\nocite{Gambino:2017vkx}

Our method studies how a heavy-light meson mass depends on the mass of its heavy valence
(anti)quark~\cite{Kronfeld:2000gk,Freeland:2006nd,Gambino:2017vkx}.
Like the quarkonium correlators, our approach requires only continuum perturbation theory.
On the other hand, the binding energy of a heavy-light meson is of order \LamQCD, so it is necessary to use heavy quark effective
theory (HQET) to separate long- and short-distance scales.
In this way, we can obtain the masses of the charm and bottom quarks and, at the same time, HQET matrix
elements~\cite{Kronfeld:2000gk}.
Because this analysis uses as inputs the bare masses of the up, down, and strange quarks---tuned to reproduce the pion and kaon
masses~\cite{Bazavov:2017fBD}, it also yields the renormalized masses of these quarks.
 
Following Ref.~\cite{Kronfeld:2000gk}, our analysis is based on the HQET formula for the heavy-light meson mass~\cite{Falk:1992wt}
\begin{equation}
    M_{H^{(*)}} = m_h + \Lambdabar + \frac{\mu_\pi^2}{2m_h} - d_{H^{(*)}} \frac{\mu_G^2(m_h)}{2m_h} + \order\left( m_h^{-2} \right),
   \label{eq:mQ_2_MH}
\end{equation}
where $M_{H^{(*)}}$ is the pseudoscalar (vector) meson mass; $m_h$ is the heavy-quark mass; and $\Lambdabar$,
$\mu_\pi^2$, and $\mu_G^2(m_h)$ are matrix elements of HQET operators with dimension~4 and~5.
The last three correspond to the energy of the light quarks and gluons, the heavy quark's kinetic energy, and the spin-dependent
chromomagnetic energy, with coefficient $d_H=1$ for pseudoscalar mesons and $d_{H^*}=-\third$ for vector mesons.
The chromomagnetic operator has an anomalous dimension, known to three loops~\cite{Grozin:2007fh}, so $\mu_G^2(m_h)$ depends
logarithmically on the mass~$m_h$.
The strategy is to use lattice QCD to compute $M_{H^{(*)}}$ as a function of~$m_h$ and fit Eq.~(\ref{eq:mQ_2_MH}) to distinguish
the terms on the right-hand side including, in principle, higher orders in $1/m_h$~\cite{Kronfeld:2000gk}.

The utility of Eq.~(\ref{eq:mQ_2_MH}) rests on the definition of the quark mass $m_h$.
In HQET, the natural definition is the pole mass (also known as the on-shell mass).
Although the pole mass is infrared finite~\cite{Kronfeld:1998di} and gauge independent~\cite{Kronfeld:1998di,Breckenridge:1994gs} at
every order in perturbation theory, its value is ambiguous when all orders are considered~\cite{Bigi:1994em,Beneke:1994sw}.
At large orders, the coefficients of the self energy grow factorially, and a possible interpretation via Borel summation is
obstructed by a series of renormalon singularities~\cite{Bigi:1994em,Beneke:1994sw}.
This behavior is a manifestation of the strongly-coupled long-range gluon field that, remarkably, appears in perturbation theory.
Note that because $M_H$ is unambiguous, the ambiguity in $m_h$ must be canceled by those in $\Lambdabar$, $\mu_\pi^2$, and
higher-dimension terms.%
\footnote{By forming the spin average, $\quarter(M_H+3M_{H^*})$, and spin difference, $M_{H^*}-M_H$, it is easy to see that 
spin-independent and spin-dependent ambiguities are distinct.}

To address this problem, some of us introduced the minimal renormalon-subtracted (MRS) mass in a companion
paper~\cite{Brambilla:2017mrs}.
It is defined by Eq.~(2.24) of Ref.~\cite{Brambilla:2017mrs},
\begin{equation}
    m_\MRS = \mbar\left(1+\sum_{n=0}^{\infty} \left[r_n-R_n\right] \alpha_s^{n+1}(\mbar) 
        + J_\MRS(\mbar) \right),
    \label{eq:mMRS}
\end{equation}
where $\mbar=m_\MSbar(m_\MSbar)$, the $r_n$ are the coefficients relating the \MSbar\ mass to the pole mass, $R_n$ denote their
asymptotic behavior, and $J_\MRS(\mbar)$, which is defined in Eqs.~(2.25) and~(2.26) of Ref.~\cite{Brambilla:2017mrs}, is the
unambiguous part of the Borel sum of $\sum R_n\alpha_s^{n+1}$.
To compute $m_\MRS$ one uses the known behavior of the $R_n$~\cite{Beneke:1994rs,Pineda:2001zq,Komijani:2017vep}, including their
overall normalization~\cite{Komijani:2017vep}.
In deriving Eq.~(\ref{eq:mMRS}), \rcite{Brambilla:2017mrs} puts the leading renormalon ambiguity into a specific quantity of
order~$\Lambda_\MSbar$, denoted $\delta m$, and transfers it from $m_h$ to~\Lambdabar.
Below we write $m_{h,\MRS}$ and $\Lambdabar_\MRS$ to denote the unambiguous definitions of $m_h$ and \Lambdabar\ in the MRS scheme.

A second feature of our technique may seem almost trivial.
In Eq.~(4.7), Ref.~\cite{Brambilla:2017mrs} rewrites $m_{h,\MRS}$ as
\begin{subequations}
    \label{eq:rabbit}
    \begin{align}
        m_{h,\MRS} &= \frac{m_{r,\MSbar}(\mu)\,am_h}{m_{h,\MSbar}(\mu)\,am_r} m_{h,\MRS} 
        \label{eq:rabbit-a} \\
            &= m_{r,\MSbar}(\mu) \frac{\mbar_h}{m_{h,\MSbar}(\mu)} \frac{m_{h,\MRS}}{\mbar_h} \frac{am_h}{am_r},
        \label{eq:rabbit-b}
    \end{align}
\end{subequations}
where $am_r$ is the bare mass (in lattice units) of staggered fermions, and the subscript~$r$ labels a reference mass; see
Sec.~\ref{sec:HQET-SChPT_function}.
Owing to the remnant chiral symmetry of staggered fermions, the first factor in Eq.~(\ref{eq:rabbit-a}) is $1+\order(a^2)$.
In Eq.~(\ref{eq:rabbit-b}), the factors are, respectively, a convenient fit parameter, the factor to run from scale $\mu$ to
$\mbar_h$, the quantity in the big parentheses in Eq.~(\ref{eq:mMRS}), and the ratio of the freely chosen heavy-quark lattice mass
to the reference mass.
Equation~(\ref{eq:rabbit}) plays a key role: with $r_n$ for \MSbar\ in Eq.~(\ref{eq:mMRS}), the first factor in
Eq.~(\ref{eq:rabbit-b}) is in the \MSbar\ scheme; with $J_\MRS$ removing the leading renormalon ambiguity, the product on the
right-hand side of Eq.~(\ref{eq:rabbit-b}) is indeed the MRS mass.
By taking $m_r=0.4m_s$ (the so-called $p4s$ approach), the analysis yields $m_s$ as well as the heavy-quark masses $m_c$ and $m_b$.

The third important feature of our work is a data set with a wide range of lattice spacing, heavy-quark mass, and light valence and
sea masses.
These data, which were generated in a companion project to compute the $B$- and $D$-meson decay constants~\cite{Bazavov:2017fBD},
are very precise, with statistical errors of 0.005--0.12\%.
It is very challenging to take advantage of the statistical power and parameter range of the data set.
In this paper, we use heavy-meson rooted all-staggered chiral perturbation theory~\cite{Bernard:2013qwa} (\aschpt) to describe the
dependence of the heavy-light pseudoscalar meson masses on the light mesons.
To make possible a fit to lattice data, \rcite{Brambilla:2017mrs} combined the next-to-leading-order \aschpt\ with the MRS mass to
write heavy-light meson masses as a function of lattice spacing and heavy- and light-quark masses.
The fit function, by construction, has the correct nonanalytic form in the chiral and HQET limits.
Here, it is extended with enough analytic terms to mimic higher-order corrections and obtain a good fit.

We use 24 ensembles generated by the MILC Collaboration~\cite{Bazavov:2010ru,\rHISQrCONFIGS,milc_hisq} with four flavors of
sea quarks using the highly-improved staggered quark (HISQ) action~\cite{Follana:2006rc} and a one-loop~\cite{Hart:2008sq}
tadpole-improved~\cite{Alford:1995hw} Symanzik-improved gauge
action~\cite{Weisz:1982zw,Weisz:1983bn,Curci:1983an,Luscher:1984xn,Luscher:1985zq}.
The algorithm for the quark determinant uses the fourth-root procedure to remove the unwanted taste degrees of freedom~%
\cite{Marinari:1981qf,Follana:2004sz,Durr:2004as,Durr:2004ta,Wong:2004nk,Shamir:2004zc,Prelovsek:2005rf,Bernard:2006zw,Durr:2006ze,%
Bernard:2006ee,Shamir:2006nj,Bernard:2007qf,Kronfeld:2007ek,Donald:2011if}.
A thorough description of the simulation program can be found in Ref.~\cite{\rHISQrCONFIGS}.
Since then, the simulations have been extended to smaller lattice spacings; up-to-date details are in Ref.~\cite{Bazavov:2017fBD}.
Our procedures for calculating pseudoscalar meson correlators and for finding masses and amplitudes from these correlators are
described in Refs.~\cite{\rFD2014,Bazavov:2017fBD}.
The amplitudes are used in Ref.~\cite{Bazavov:2017fBD} to calculate the decay constants of $B$ and $D$ mesons, and the corresponding
meson masses are used here.

A preliminary report of this analysis can be found in Ref.~\cite{Komijani:2016jrh}.
Instead of the MRS mass, at that time we used the renormalon-subtracted (RS) mass~\cite{Pineda:2001zq}, which also subtracts the
leading renormalon ambiguity but at the same time introduces a factorization scale~$\nu_f$.
In principle, the \MSbar\ masses emerging from Eqs.~(\ref{eq:rabbit}) and~(\ref{eq:mQ_2_MH}) should not depend on $\nu_f$, but we
found more dependence than one would like.
Moreover, it turns out to be necessary to introduce three scales in all, $\nu_f<\mu<m_h$, with $\mu$ being used for
$\alpha_s$~\cite{Komijani:2016jrh}.
For that reason, we prefer the MRS over the RS mass.
 
This paper is organized as follows.
\Secref{sims} contains a description of the lattice-QCD simulations, focusing on the way we eliminate the
lattice scale in favor of physical units.
In \secref{HQET-SChPT_function}, we present our function of quark masses and lattice spacing that describes masses of heavy-light
pseudoscalar mesons.
In \secref{HQET-SChPT_fit}, we perform a combined-correlated fit to the meson masses; the fit is then extrapolated to the continuum
and interpolated to physical values of the light quark masses.
In \secref{Results}, we present our final results for the masses of the strange, charm and bottom quarks as well as quark mass
ratios $m_c/m_s$, $m_b/m_s$, and $m_b/m_c$.
Combining our results with our separate determination of the quark-mass ratios $m_u/m_d$ and $m_s/m_l$, where
$m_l\equiv\frac{1}{2}(m_u+m_d)$, we also report the up- and down-quark masses.
In addition, we present our lattice-QCD determinations of $\Lambdabar_\MRS$, $\mu_\pi^2$, and
$\mu_G^2(m_b)$ as well as flavor splittings and low-energy constants of heavy-meson chiral perturbation theory.
Section~\ref{sec:Outlook} compares our main results with work in the literature and offers some remarks on further work.
An appendix gives the correlation matrices of the MRS masses of the charm and bottom quarks with HQET matrix elements,
and of the charm-quark mass and quark-mass ratios.

\section{Simulations summarized}
\label{sec:sims}

The lattice data used in this work come from the same correlation functions used to
determine leptonic decay constants of charmed and $b$-flavored mesons in a companion paper~\cite{Bazavov:2017fBD}.
For a full description of the simulation, the reader should consult Ref.~\cite{Bazavov:2017fBD}.
Here we provide a brief summary.

We employ a data set that includes ensembles with five values of lattice spacings ranging from approximately 0.12~fm to 0.03~fm,
enabling good control over the continuum extrapolation.
Ensembles at a sixth lattice spacing, approximately 0.15~fm, are used only to estimate the continuum
extrapolation error.
The data set includes ensembles with the light (up-down), strange, and charm sea masses close to their physical values
(``physical-mass ensembles'') at all but the smallest lattice spacing, 0.03~fm.
The data set also includes ensembles where either the mass of light sea quarks is heavier than in nature, or the mass of the strange
sea quark is lighter than in nature, or both.
As in \rcite{Bazavov:2017fBD}, we set the scale of the lattice spacing $a$ with a two-step procedure that uses the value of $f_\pi$
from the Particle Data Group (PDG), $\fpiPDG=130.50(13)$~MeV~\cite{\rPDG2016}, combined with the so-called $p4s$ method.

The first step in the scale-setting procedure takes \fpiPDG\ to set the overall scale on each \emph{physical-mass} ensemble.
On these ensembles, we tune the valence light, strange, and charmed quark masses to reproduce the pion, kaon, and $D_s$-meson masses.
Then we calculate $M_{p4s}$ and $f_{p4s}$, which are the mass and decay constant of a pseudoscalar meson with both valence-quark
masses set equal to $m_{p4s}\equiv0.4m_s$.
We then form the ratio $R_{p4s}\equiv f_{p4s}/M_{p4s}$ and take the continuum limit of $f_{p4s}$ and $R_{p4s}$.
These values and those of the quark mass ratios are then used as inputs to the second step of the procedure, which we call the
$p4s$~method.
In the $p4s$ method, the values of $am_{p4s}$ and $af_{p4s}$ are calculated on a given physical-mass ensemble, with $a\neq0$, by
adjusting the valence-quark mass until $af_{p4s}/aM_{p4s}$ equals the physical-mass continuum limit of~$R_{p4s}$:
\begin{equation}
    R_{p4s}(m_{p4s};\{\half(m_u+m_d),m_s,m_c\};a) = R_{p4s}(0.4m_{s};\{\half(m_u+m_d),m_s,m_c\};0).
\end{equation}
In the $p4s$ method, all ensembles at the same bare gauge coupling, $\beta=10/g_0^2$, as a given physical-mass ensemble are chosen
to have the same lattice spacing $a$ and the same $am_{p4s}$.
This choice is known as a \emph{mass-independent scale-setting scheme}.

At $a\approx0.03$~fm, we have only a 0.2$m_s$ ensemble, so this procedure cannot be carried out.
In this case, we rely on the derivatives with respect to $a$, which are given in Ref.~\cite{Bazavov:2014wgs,*Bazavov:2014lja}.

\section{Construction of the fit function}
\label{sec:HQET-SChPT_function}

In this section, we discuss in detail how to construct a function of quark masses and lattice spacing that describes masses of
heavy-light pseudoscalar mesons.
To this end, we use three effective field theories (EFTs), HQET, and \aschpt, as mentioned already, and the Symanzik effective theory
of cutoff effects~\cite{Symanzik:1980rcnt,Symanzik:1983dc,Luscher:1984xn}.
We start with the merger of HQET and \aschpt~\cite{Brambilla:2017mrs} 
and incorporate generic lattice-spacing dependence, as well as higher-order terms in HQET and \aschpt.
Putting everything together, we obtain an EFT fit function for masses of heavy-light pseudoscalar mesons.

\subsection{\boldmath Leading-order \texorpdfstring{\chpt}{ChPT}}
\label{sec:SChPT}

Let us start with fixing our notation for quark masses associated with lattice ensembles with $2{+}1{+}1$~flavors of quarks.
We use $m'_l$, $m'_s$, and $m'_c$ to denote the simulation masses of the light (up-down), strange, and charm quarks, respectively;
without the primes, we use $m_l = \frac{1}{2}(m_u+m_d)$, $m_s$, and $m_c$ to denote the correctly tuned masses of the corresponding
quarks; last, we use $m_q$ to denote a generic light quark mass.
Further, we use $H_x^{(*)}$ to denote a generic heavy-light pseudoscalar (vector) meson composed of a light valence quark $x$ and a
heavy valence antiquark $\bar{h}$.
We also use $m_{h,\MRS}$, $m_{h,\MSbar}$, and $am_h$ to denote the MRS, $\MSbar$, and bare masses of antiquark $\bar{h}$,
respectively.
The relations between $m_{h,\MRS}$, $m_{h,\MSbar}$, and $am_h$ are discussed in \secref{MRS-bare-mass}.

In \aschpt, the mass of $H_x^{(*)}$ meson is described by Eq.~(4.2) of Ref.~\cite{Brambilla:2017mrs}
\begin{align}
    M_{H_x^{(*)}}(m_x;\{m'_l,m'_l,m'_s\};a) &= m_{h,\MRS} + \Lambdabar_\MRS + \frac{\mu_\pi^2
        - d_{H^{(*)}}\mu_G^2(m_h)}{2m_{h,\MRS}} + 2\lambda_1 B_0 m_x
    \label{eq:M_Hv} \\ &
        + 2\lambda'_1 B_0(2m'_l + m'_s) + \delta M_{H_x^{(*)}}(m_x;\{m'_l,m'_l,m'_s\};a) - \cC^{(*)} ,
    \nonumber
\end{align}
where $B_0$ is the low energy constant (LEC) in the relation $m^2_\pi = B_0(m_u+m_d)$ between the pion mass and the quark mass;
$d_{H^{(*)}}=1$ ($-\third$) for pseudoscalar (vector) mesons; $\lambda_1$ and $\lambda'_1$ are LECs that appear in (continuum)
heavy-meson chiral perturbation theory (\hmchpt)~\cite{Boyd:1994pa}; and $\delta M_{H_x^{(*)}}$ is the one-loop corrections to the
mass of the $H_x^{(*)}$ meson in \aschpt~\cite{Brambilla:2017mrs}.
The arguments of $M_{H_x^{(*)}}$ and $\delta M_{H_x^{(*)}}$ in \eq{M_Hv} correspond to the light valence-quark mass; the set of
three light sea-quark masses, which are not necessarily tuned to their physical values; and the lattice spacing~$a$.
As usual for a one-loop \chpt\ result, $\delta M_{H_x}$ contains a term nonanalytic as $m_\pi^2\to0$ (a ``chiral log'').
For the pseudoscalar mesons with $(2+1)$ light flavors in the sea, we have.
\begin{align}
    \delta M_{H_x} &= - \frac{3g_\pi^2}{16\pi^2f^2} \Biggl\{
        \frac{1}{16}\sum_{\mathscr{S},\Xi'} K_1(m_{\mathscr{S}x_{\Xi'}},\Delta^* + \delta_{\mathscr{S}x})
    \label{eq:chiral-form:2p1} \\
        & \qquad + \frac{1}{3}\sum_{j\in \cM_I^{(2,x)}} \frac{\partial}{\partial m^2_{X_I}}
            \left[R^{[2,2]}_{j}(\cM_I^{(2,x)};     \mu^{(2)}_I) K_1(m_{j},   \Delta^*) \right]
    \nonumber \\
        & \qquad + \biggl( a^2\delta'_V \sum_{j\in \hat{\cM}_V^{(3,x)}} \frac{\partial}{\partial m^2_{X_V}}
            \left[R^{[3,2]}_{j}(\hat{\cM}_V^{(3,x)}; \mu^{(2)}_V) K_1(m_{j},   \Delta^*) \right] + [V\to A]\biggr)  
	    \Biggr\}
    \nonumber \\
        & \qquad +  a^2 \frac{3g_\pi^2}{16\pi^2f^2} 
        \left[ \lambda'_{a^2} \bar{\Delta} \sum_\mathscr{S}\delta_{\mathscr{S}x} 
        + \lambda_{a^2} \Delta^* \left( 3\bar{\Delta} -\third\Delta_I + \delta'_V + \delta'_A\right)\right].
    \nonumber
\end{align}
where the indices $\mathscr{S}$ and $\Xi$ run over light sea-quark flavors and meson tastes, respectively; $M_{\mathscr{S}x,\Xi}$
is the mass of the pseudoscalar meson with taste $\Xi$ and flavors $\mathscr{S}$ and $x$; $\Delta^*$ is the lowest-order hyperfine
splitting; $\delta_{\mathscr{S}\vt}$ is the flavor splitting between a heavy-light meson with light quark of flavor
$\mathscr{S}$ and one of flavor $\vt$; $g_\pi$ is the $H$-$H^*$-$\pi$ coupling; $\delta'_A$ and $\delta'_V$ are the
taste-breaking hairpin parameters; $a^2 \bar\Delta$ is the mean-squared pion taste splitting;
and $\lambda_{a^2}$ and $\lambda'_{a^2}$ are parameters in \schpt\, related to taste breaking in meson masses.
Definitions of the residue functions $R_j^{[n,k]}$, the sets of masses in the residues, and the chiral-log function $K_1$ at
infinite and finite volumes are given in \rcite{Brambilla:2017mrs} and references therein.
The expression for $\delta M_{H^*_x}$ is also given in \rcite{Brambilla:2017mrs}, but because we have lattice data only for 
pseudoscalar mesons, it is not needed here.

In \eq{M_Hv}, we set
\begin{equation}
    \cC^{(*)} = 2\lambda_1 B_0 m_q + 2\lambda'_1 B_0(2m_l + m_s) + \delta M_{H_q^{(*)}}(m_q; \{m_l,m_l,m_s\}; 0)
    \label{eq:fit-function:C}
\end{equation}
so that in the continuum limit the usual expression
\begin{equation}
    M_{H_q^{(*)}}(m_q;\{m_l,m_l,m_s\}; 0) = m_{h,\MRS} + \Lambdabar_\MRS + \frac{\mu_\pi^2
        - d_{H^{(*)}}\mu_G^2(m_h)}{2m_{h,\MRS}}
    \label{eq:M_Hl}
\end{equation}
is recovered for physical values of sea-quark masses and $m_x=m_q$.
With this choice for $\cC$, the values that we obtain for $\Lambdabar_\MRS$, $\mu_\pi^2$ and $\mu_G^2(m_h)$ are readily applicable
for calculations in HQET.%
\footnote{Note that in the context of Eq.~(\ref{eq:M_Hl}), the matrix elements $\Lambdabar_\MRS$, $\mu_\pi^2$ and 
$\mu_G^2(m_h)$ depend on the light-quark masses.}
In this work, we set $m_q=\half(m_u+m_d)$, and we report $\Lambdabar_\MRS$, $\mu_\pi^2$ and $\mu_G^2(m_h)$ for this choice.

At this stage, the fit parameters are $m_{r,\MSbar}(\mu=2~\GeV)$ via Eq.~(\ref{eq:rabbit}), $\Lambdabar_\MRS$, the kinetic energy
$\mu_\pi^2$, the chromomagnetic energy $\mu_G^2(m_b)$ from which we obtain $\mu^2_G(m_h)$ as in \eq{mu_G_sq(mh)} below,
and the LECs $\lambda_1$, $\lambda'_1$, $\lambda_{a^2}$, and $\lambda'_{a^2}$.
Ideally, one would have data for both pseudoscalar- and vector-meson masses, and then one could set up
separate fits for spin-independent and spin-dependent terms.
In this work, however, only the pseudoscalar masses are available.
The experimental masses of the $B^*$ and $B$ mesons can be used to estimate
\begin{equation}
    \mu_G^2(m_b) \approx \frac{3}{4}(M_{B^*}^2 - M_B^2) = 0.36~\GeV^2,
    \label{eq:mu_G_sq(mb)}
\end{equation}
which neglects contributions to the hyperfine splitting suppressed by a power of $1/m_b$.
The chromomagnetic operator has an anomalous dimension, however, so we obtain
$\mu_G^2(m_h)$ in Eq.~(\ref{eq:M_Hv}) with
\begin{equation}
    \mu_G^2(m_h) = \frac{C_\text{cm}(m_h)}{C_\text{cm}(m_b)} \mu_G^2(m_b),
    \label{eq:mu_G_sq(mh)}
\end{equation}
using the three-loop relation~\cite{Grozin:2007fh} for the Wilson coefficient~$C_\text{cm}(m_h)$.
For four active flavors,
\begin{equation}
    C_\text{cm}(m_h) = \alpha_s^{9/25} \left(1 + 0.672355 \alpha_s + 1.284 \alpha_s^2\right),
    \label{eq:C_cm}
\end{equation}
where $\alpha_s = \alpha_{\MSbar}(\mbar_h)$.

As discussed in Sec.~\ref{sec:Introduction} and \rcite{Brambilla:2017mrs}, the matrix elements of HQET suffer in general
from ambiguities related to renormalon singularities, although the ambiguities cancel in observables such as the meson mass.
For instance, the ambiguity in $\Lambdabar$ cancels the leading-renormalon ambiguity in the pole mass.
By construction, only the leading renormalon is removed to define the MRS mass.
In principle, renormalon ambiguities in $\mu_\pi^2$ and $\mu_G^2(m_h)$ remain.
In practice, numerical investigation indicates that the subleading infrared renormalon of the pole mass is
small~\cite{Brambilla:2017mrs}, which implies that the corresponding renormalon ambiguity in $\mu_\pi^2$ is not large.
Moreover, the leading spin-dependent renormalon in $\mu_G^2$ is suppressed by a further power of~$1/m_h$.

\subsection{Higher-order terms in \boldmath\texorpdfstring{\chpt}{ChPT}}
\label{sec:Higher-corrections}

Because we have very precise data with statistical errors of 0.005--0.12\%, we can
anticipate that NLO \chpt\ is not enough to fully describe the quark-mass dependence, especially for data with $m_x$ near~$m_s$.
We therefore extend the function given in \eq{M_Hv} by adding higher-order analytic corrections in powers of light quark masses and
in inverse powers of the heavy quark mass.
For the expansion in inverse powers of the heavy-quark mass, we introduce the dimensionless variable
\begin{equation}
    w_h = \frac{\LamHQET}{m_{h,\MRS}},
    \label{eq:minv}
\end{equation}
with $\LamHQET=600~\MeV$.
Then the natural size of coefficients of the $1/m_h$ corrections is of order~1.
For expansion in light quark masses, following Refs.~\cite{\rFD2014,Bazavov:2017fBD}, we define dimensionless quark masses, which
are natural expansion parameters in \chpt:
\begin{equation}
    x_q  \equiv \frac{B_0}{4 \pi^2 f_{\pi}^2} m_q,
    \eqn{x-defs}
\end{equation}
where $q$ can be either the valence or sea light quarks.
For simplicity, we drop the primes on the simulation $x_q$s.
The quark masses in the formula for $\delta M_{H_x}$ can also be expressed in terms of $\{x_x, x_l, x_s\}$.

We include all mass-dependent analytic terms at order $x_q^2$ by adding
\begin{equation}
    f_\pi\left[q_1 x_x^2 + q_2x_x(2x_l + x_s) + q_3 (2x_l+x_s)^2 + q_4(2x_l^2+x_s^2)\right]
    \label{eq:NNLO-analytic}
\end{equation}
to the expression for $M_{H_x}$ in \eq{M_Hv}.
With $f_\pi$ to set the overall scale of these higher-order terms,
the coefficients $q_i$ become of order~1 or less.
We also include all mass-dependent analytic terms at order~$x_q^3$, namely
\begin{equation}
    x_x^3,\; x_x^2(2x_l + x_s),\; x_x(2x_l+x_s)^2,\; x_x (2x_l^2+x_s^2),\;
        (2x_l + x_s)^3,\; (2x_l+x_s)(2x_l^2+x_s^2),\; 2x_l^3+x_s^3 .
    \label{eq:NNNLO-analytic}
\end{equation}
In practice, one can expect the terms without $x_x$ to be less important, but we keep all of them for consistent power counting.

To improve the expansion in inverse powers of the heavy quark, we add
\begin{equation}
    \LamHQET \left(\rho_1 w_h^2 + \rho_2 w_h^3 + \rho_3 w_h^4\right)
    \label{eq:rho-kappa}
\end{equation}
with three fit parameters $\rho_i$ to the right-hand side of \eq{M_Hv}.
We also add $w_h$ and $w_h^2$ corrections to the LECs $\lambda_1$, $\lambda'_1$ and $g_\pi$; and $w_h$ corrections to the fit
parameters $q_i$ in \eq{NNLO-analytic}.

The heavy quark mass also affects the hyperfine splitting $\Delta^*$ and the flavor splitting~$\delta_{\mathscr{S}x}$ in
\eq{chiral-form:2p1}.
Although we could express these quantities in terms of $\mu_G^2(m_h)$ and $\lambda_1$, we exploit
the experimental values for the hyperfine splittings and flavor splittings in the $D$ and $B$ systems
to calculate $\Delta^*$ and $\delta_{\mathscr{S}x}$ for different quark masses.
See our companion paper on decay constants~\cite{Bazavov:2017fBD} for details.

We now discuss the effects of mistuning in the sea charm-quark mass $m'_c$.
The effects can be divided into two parts: the effects on the pole mass (and, hence, the MRS mass) and the effects on the effective
theory after the charm quark is integrated out.
The former effects are taken into account in calculating the MRS mass from the $\MSbar$ mass; cf.\ Eq.~(\ref{eq:mSI2mMRS}).
We treat the latter effects as in Ref.~\cite{Bazavov:2017fBD}.
We use $\LamQCD^{(3)}(m'_c)$ to denote the effective value of $\LamQCD$ when the charm quark with mass $m'_c$ is integrated out.
At leading order in weak-coupling perturbation theory, one obtains \cite[Eq.~(1.114)]{Manohar:2000dt}
\begin{equation}
    \frac{\LamQCD^{(3)}(m'_c)}{\LamQCD^{(3)}(m_c)} = \left(\frac{m'_c}{m_c}\right)^{2/27},
    \label{eq:mc:mc}
\end{equation}
where $m_c$ is the correctly tuned value of charm-quark mass.
Assuming $m'_c\approx m_c$, we take the effects of the mistuned mass $m'_c$ into account by multiplying $\Lambdabar_\MRS$ with
\begin{equation}
    \left(\frac{m'_c}{m_c}\right)^{2/27}\left(1 + \frac{2k'_1}{27}\frac{m'_c-m_c}{m'_c}\right),
\end{equation}
where the extra fit parameter~$k'_1$ describes higher-order corrections to Eq.~(\ref{eq:mc:mc}).

We must also include generic lattice artifacts in our analysis.
Taste-breaking discretization errors from staggered fermions are already included in~\eq{chiral-form:2p1}.
In addition to these effects, various discretization errors, from gluons for example, must be taken into account.
We include the leading lattice artifacts for $\Lambdabar_\MRS$ by replacing
\begin{equation}
    \Lambdabar_\MRS \to \Lambdabar_\MRS \left[ 1 + \bar{c}_1 \alpha_s (a\Lambda)^2 + \bar{c}_2 (a\Lambda)^4 \right],
\end{equation}
where $\Lambda$ is the scale of generic discretization effects, set to 600~MeV in this analysis.
The factor of $\alpha_s$ in the second-order term arises because
the HISQ action is tree-level improved to order~$a^2$.
Note that $\Lambdabar_\MRS$ is not affected by heavy-quark discretization errors.
As discussed in the Appendices of Ref.~\cite{Bazavov:2017fBD}, at leading order (LO) in HQET, heavy-quark
discretization errors only affect the normalization of the heavy-quark state.
Thus, $\Lambdabar_\MRS$ and also $\lambda_1$, $\lambda'_1$ and $g_\pi$ at leading order in $1/m_h$ are free of heavy-quark
discretization errors.
For $\lambda_1$ we replace
\begin{equation}
    \lambda_1 \to \lambda_1\left[ 1 + c_1 \alpha_s (a\Lambda)^2 + c_2 (a\Lambda)^4 +
        c_3 w_h \alpha_s (am_h)^2\right],
\end{equation}
where the $c_3$ term is added to incorporate effects of heavy-quark discretization errors.
We incorporate similar corrections for $\lambda'_1$ and $g_\pi$.
Finally, we add $\alpha_s (a\Lambda)^2$ and $\alpha_s (am_h)^2$ corrections to
$\mu_\pi^2$ and $\mu_G^2(m_b)$, and $\alpha_s (a\Lambda)^2$ corrections to the parameters $q_i$ in \eq{NNLO-analytic}.

\subsection{Heavy-quark mass}
\label{sec:MRS-bare-mass}
Although the MRS mass is the key to our interpretation of the HQET mass formula, as indicated in Eq.~(\ref{eq:rabbit}) we arrange 
the fit to yield the \MSbar~mass.
For $am\ll1$, the relation between the \MSbar\ and bare masses is
\begin{equation}
    m_\MSbar(\mu) = \frac{am}{a} \left\{1 + \alpha_s\left[-(2/\pi)\log(a \mu)+ k_0 + k_1 (am)^2 + \cdots \right]
        + \order(\alpha_s^2)\right\}, \label{eq:bare2MSbar}
\end{equation}
where $a$ in the denominator is set from the scale setting quantity (here $f_{p4s}$, as described in Sec.~\ref{sec:sims}).
With staggered fermions, there is no additive mass renormalization,
and to eliminate tree-level discretization errors from \eq{bare2MSbar}, we take the mass $am$ to be the tree-level pole mass.%
\footnote{The exact relation between $m_0$ and $m$ can be found in Appendix~A of Ref.~\cite{Bazavov:2017fBD}.}
Taking the ratio between two masses%
\footnote{For Wilson fermions with order-$a$ improvement, the following arguments hold for the mass defined through the axial 
Ward identity, apart from details about the lattice artifacts.}
\begin{equation}
    \frac{m_{h,\MSbar}(\mu)}{m_{r,\MSbar}(\mu)} = \frac{am_h}{am_r} \left( 1 + \alpha_\MSbar(\mu) 
        \left\{k_1\left[(a m_h)^2 - (a m_r)^2\right] + \cdots \right\} + \cdots \right) , 
  \label{eq:bare_2_MS} 
\end{equation}
where the dots stand for higher-order terms in $a^2$ and $\alpha_s$.
In fact, each higher order in $\alpha_s$ is also multiplied by a quantity of order~$a^2$, as stated in the Introduction.
In this analysis, we set the reference-quark mass $m_r$ to $m_{p4s}\equiv 0.4m_s$ and the scale of the $\MSbar$ scheme to
$\mu=2$~GeV.
Thus, $m_{p4s,\MSbar}(2~\text{GeV})$ is a free parameter left to be determined in the fit to lattice data; cf.\
Eq.~(\ref{eq:rabbit-b}).

To incorporate further heavy-quark discretization effects into \eq{bare_2_MS},
we multiply the right-hand side of Eq.~(\ref{eq:rabbit-b}) by
\begin{equation}
    \left[ 1 + \alpha_\MSbar(2~\GeV) \sum_{n=1}^{4} k_n x_{h}^{n} \right] ,
    \label{eq:fit-form:bare2MS}
\end{equation}
where the dimensionless coefficients $k_n$ are free fit parameters, and
\begin{equation}
    x_{h} = \left(2am_h/\pi\right)^2 - \left(2am_{p4s}/\pi\right)^2 
        \approx \left(2am_h/\pi\right)^2  .
\end{equation}
We multiply $am_h$ by a factor of $2/\pi$ so that the parameters $k_n$ become of order~1, based on the radius of convergence of
various tree-level formulas for the HISQ action; see Appendix~A of \rcite{Bazavov:2017fBD}.
Because $(am_{p4s}/am_c)^2\approx0.001$, the effects of a nonzero value of $am_{p4s}$ are negligible compared with the heavy-quark
discretization effects.
To incorporate generic lattice-spacing dependence into our analysis, we additionally multiply the 
right-hand side of Eq.~(\ref{eq:rabbit-b}) by
\begin{equation}
  \left[ 1 + \tilde{c}_1 \alpha_s (a\Lambda)^2 + \tilde{c}_2 (a\Lambda)^4 +  \tilde{c}_3(a\Lambda)^6 \right] .
  \label{eq:m_p4s_2GeV:lattice-artifacts}
\end{equation}

To complete our approach to introducing $m_{p4s,\MSbar}(2~\GeV)$ via $m_{h,\MRS}$, we must describe the calculation of the second
and third factors in Eq.~(\ref{eq:rabbit-b}).
The second factor simply uses the anomalous dimension to run from $\mu=2~\GeV$ to the self-consistent scale $\mbar_h\equiv
m_{h,\MSbar}(\mbar_h)$
\begin{equation}
    \frac{\mhbar}{m_{h,\MSbar}(\mu)} = \frac{C\left(\alpha_\MSbar(\mhbar)\right)}{C\left(\alpha_\MSbar(\mu)\right)},
    \label{eq:mMSbar2mRGI}
\end{equation}
where with four active flavors~\cite{Baikov:2014qja} 
\begin{equation}
    C(\pi u) = u^{12/25} \left[1 + 1.01413  u + 1.38921 u^2 + 1.09054 u^3  + 5.8304 u^4 + 
        \order(u^5)\right].
    \label{eq:mMSbar2mRGI_c}
\end{equation}
The coefficient of $u^4$ is obtained from the five-loop results for the quark-mass anomalous
dimension~\cite{Baikov:2014qja} and beta function~\cite{Baikov:2016tgj}.
Finally, the third factor in Eq.~(\ref{eq:rabbit-b}) is simply the relation
derived in Ref.~\cite{Brambilla:2017mrs}, which at the four-loop level reads
\begin{equation}
    \frac{m_{h,\MRS}}{\mbar_h} =
        1 + \sum_{n=0}^{3} \left[r_n-R_n\right] \alpha_s^{n+1}(\mbar_h)
        + J_\MRS(\mbar_h) + \frac{\Delta m_{(c)}}{\mbar_h} + \order(\alpha_s^5),
  \label{eq:mSI2mMRS}
\end{equation}
where the $r_n$ are known through order $\alpha_s^4$~\cite{Marquard:2015qpa,Marquard:2016dcn}; the $R_n$ depend only on the
coefficients of the beta function~\cite{Beneke:1994rs,Komijani:2017vep} up to an overall normalization, which is given in
\rcite{Komijani:2017vep}; the function $\mbar J_\MRS(\mbar)=\cJ_\MRS(\mbar)$ appears in the definition of the MRS
mass~\cite{Brambilla:2017mrs}; and~$\Delta m_{(c)}$ contains the contribution from the charm sea quark.
Because the nonzero mass of the charmed sea quark cuts off the infrared region that is the origin of factorial growth in the
$r_n$~\cite{Ball:1995ni}, we subtract the renormalon with three massless active quarks and lump the charmed loops'
contributions into~$\Delta m_{(c)}$~\cite{Ayala:2014yxa}.

The detailed formulas for $\cJ_\MRS(\mbar)$ and $\Delta m_{(c)}$ can be found in Ref.~\cite{Brambilla:2017mrs}.
The crucial aspects of Eq.~(\ref{eq:mSI2mMRS}) for the fits of the next section is that the renormalon-subtracted perturbative
coefficients are small: $r_n-R_n=(-0.1106, -0.0340, 0.0966, 0.0162)$ for $n=(0, 1, 2, 3)$ and three active flavors.
The Borel resummed renormalon is computed from a function with a convergent expansion in $1/\alpha_s$.
(In fact, our implementation of one of the factors in $\cJ_\MRS$ uses the convergent expansion until it saturates to numerical 
precision.)

\subsection{Summary formulas}

In summary, we fit our data for $aM(m_h,m_x;\{m'_l,m'_l,m'_s\};a)$ to
\begin{equation}
    \left.\frac{aM}{af_{p4s}}\right|_\text{data} f_{p4s} = \cF ,
\end{equation}
where $\cF$ is the fit function and $f_{p4s}$ is in the continuum limit.
From the preceding subsections [with free fit parameters in \param{blue} (arXiv)]:
\begin{align}
    \cF = \breve{m}_{h,\MRS} &+ \breve{\Lambdabar}_\MRS
        + \frac{\breve{\mu}_\pi^2}{2m_{h,\MRS}}
        - \frac{\breve{\mu}_G^2(m_b)}{2m_{h,\MRS}} \frac{C_\text{cm}(m_h)}{C_\text{cm}(m_b)}
    \label{eq:fit:andreas} \\ &
        + 2\breve{\lambda}_1 B_0 (m_x - m_l)
        + 2\breve{\lambda}'_1 B_0 (2m'_l+m'_s - 2m_l-m_s)
    \nonumber \\ &
        + \delta M_{H_x}(m_x;\{m'_l,m'_l,m'_s\};a)
        - \delta M_{H_l}(m_l;  \{m_l, m_l, m_s \};0)
    \nonumber \\ &
        + \LamHQET\left[\param{\rho_1}w_h^2+\param{\rho_2}w_h^3 + \param{\rho_3}w_h^4 \right]
    \nonumber \\ &
        + f_\pi\left[
              \sum_{i=1}^4 \param{q_i}\left(1 + \param{q'_i}w_h+\param{\tilde{q}_i}\alpha_sy^2\right) x_i^2
            + \sum_{j=5}^{11} \param{q_j} x_j^3 \right] ,
    \nonumber
\end{align}
where $y=(a\Lambda)^2$ and $w_h=\LamHQET/m_{h,\MRS}$.
The \aschpt\ self energy $\delta M_{H_x}$ depends on $\param{f}$, $\param{\lambda_{a^2}}$, $\param{\lambda'_{a^2}}$,
$\breve{g}_\pi$, $\param{\delta'_V}$, and $\param{\delta'_A}$, as well as $\Delta^*$ and taste-independent~$\delta_{\mathscr{S}\vt}$.
The breved quantities are
\begin{subequations}
    \begin{align}
        \breve{\Lambdabar}_\MRS &= \param{\Lambdabar_\MRS} \left(1 + \param{\bar{c}_1}\alpha_s y + \param{\bar{c}_2}y^2 \right)
        \left(\frac{m'_c}{m_c}\right)^{2/27}\left(1+\param{k'_1}\frac{\delta m'_c}{m_c}\right) , \\
        \breve{\lambda}_1  &= \param{\lambda_1}  \left(1 + \param{c_1} \alpha_s y + \param{c_2} y^2 + \param{c_3} \bar{w}_h\alpha_s y
            + \param{c_4} \bar{w}_h + \param{c_5} \bar{w}_h^2 + \param{c_6} \bar{w}_h^3 \right) , \\
        \breve{\lambda}'_1 &= \param{\lambda'_1} \left(1 + \param{c'_1}\alpha_s y + \param{c'_2}y^2 + \param{c'_3}\bar{w}_h\alpha_s y
            + \param{c'_4}\bar{w}_h + \param{c'_5}\bar{w}_h^2 + \param{c'_6}\bar{w}_h^3 \right) , \\
        \breve{g}_\pi      &= \param{g_\pi}     \left(1 + \param{g_1} \alpha_s y + \param{g_2} y^2 + \param{g_3} \bar{w}_h\alpha_s y
            + \param{g_4} \bar{w}_h + \param{g_5} \bar{w}_h^2 + \param{g_5} \bar{w}_h^3 \right), \\
        \breve{\mu}_\pi^2  &= \param{\mu_\pi^2} \left(1 + \param{p_\pi} \alpha_s y + \param{r_\pi} \alpha_s x_h^2\right), \\
        \breve{\mu}_G^2(m_b) &= \param{\mu_G^2(m_b)} \left(1 + \param{p_G} \alpha_s y + \param{r_G} \alpha_s x_h^2\right),
    \end{align}
\end{subequations}
where $\bar{w}_h=w_h-\LamHQET/m_{c,\MRS}$; further
\begin{align}
    \hspace*{-1em}
    \breve{m}_{h,\MRS} = \param{m_{p4s,\MSbar}(2~\GeV)} &
        \left[\frac{C(\alpha_\MSbar(\mbar_h))}{C(\alpha_\MSbar(2~\GeV))}\right]_\text{Eq.~(\ref{eq:mMSbar2mRGI_c})}
        \left[\frac{m_{h,\MRS}}{\mbar_h}\right]_\text{\rcite{Brambilla:2017mrs}}
        \left[\frac{am_{0h}}{am_{0,p4s}}\right]_\text{sim}
    \times \label{eq:MRSbreve} \\ &
        \left(1 + \alpha_\MSbar(2~\GeV)\sum_{n=1}^4 \param{k_n}x_h^n\right) \times
        \left(1 + \param{\tilde{c}_1} \alpha_s y + \param{\tilde{c}_2} y^2 + \param{\tilde{c}_3} y^3\right). \nonumber
\end{align}
Thus, there are $\param{61}$
free parameters, 4 parameters [$\param{f}$, $\param{g_\pi}$, $\param{\mu_G^2(m_b)}$, and, in $m_{h,\MRS}/\mbar_h$,
\param{$R_0$}] with external priors, and $2$~hairpin parameters (\param{$\delta'_V$} and \param{$\delta'_A$}) from light-meson \chpt.
$\Delta^*$ and $\delta_{\mathscr{S}\vt}$ introduce 2 parameters each that are, however, frozen to reproduce PDG hyperfine and flavor
splittings. 
The total number of fit parameters is 67 (compared with 60 for the decay-constant fit~\cite{Bazavov:2017fBD}).

In Eq.~(\ref{eq:MRSbreve}), $\mbar_h$ is given self-consistently by using the formula
\begin{align}
    \mbar_h = \param{m_{p4s,\MSbar}(2~\GeV)} &
        \left[\frac{C(\alpha_\MSbar(\mbar_h))}{C(\alpha_\MSbar(2~\GeV))}\right]_\text{Eq.~(\ref{eq:mMSbar2mRGI_c})}
        \left[\frac{am_{0h}}{am_{0,p4s}}\right]_\text{sim}
    \times \label{eq:MSbarbreve} \\ &
        \left(1 + \alpha_\MSbar(2~\GeV)\sum_{n=1}^4 \param{k_n}x_h^n\right) \times
        \left(1 + \param{\tilde{c}_1} \alpha_s y + \param{\tilde{c}_2} y^2 + \param{\tilde{c}_3} y^3\right), \nonumber
\end{align}
to readjust the argument of $C(\alpha_\MSbar(\mbar_h))$.
These parameters are not new but the same as those in Eq.~(\ref{eq:MRSbreve}).

\section{EFT fit to determine the quark masses}
\label{sec:HQET-SChPT_fit}

In \secref{HQET-SChPT_function}, we have constructed a function with 67 fit parameters that is motivated by EFTs.
Here, we use this function to perform a correlated fit to partially-quenched data at five lattice spacings, from $a\approx0.12$~fm
to $\approx0.03$~fm, and at several values of the light sea-quark masses.
A~sixth lattice spacing, $a\approx0.15$~fm, is used to check discretization errors but is not included in the base fit.
At the coarsest lattice spacings, we only have data with two different values for valence heavy-quark mass: $m_h=m'_c$ and
$m_h=0.9m'_c$, where $m'_c$ is the simulation value of sea charm-quark mass in each ensemble.
It is close to but not precisely equal to the physical charm mass $m_c$ because of tuning errors.
We include data with $0.9m'_c\le m_h\le 5m'_c$ subject to the condition $am_h<0.9$, which is chosen to avoid large lattice artifacts.
For every valence heavy quark, we use several light valence quarks with masses $m_l \lesssim m_x \lesssim m_s$;
on ensembles with the mass of the strange sea quark close to its physical value,
$m_x/m'_s$ takes values in a subset of $\{0.036,0.1,0.2,0.4,0.6,1.0\}$ (in several cases the whole set).
In the base fit, we obtain the meson masses from fits to two-point correlators with three pseudoscalar states and two
opposite-parity states, which we denote ``3+2'' below.
To investigate the error arising from excited state contamination, we also use meson-mass data from (2+1)-state fits.

The values of the bare masses corresponding to the light and strange quarks are taken from combinations of the physical pion and
kaon masses, as discussed in Refs.~\cite{Bazavov:2017fBD} and~\cite{Bazavov:2014wgs}.
Similarly, the physical charmed and bottom quarks are defined so that the $D_s$- and $B_s$-meson masses take their physical values.
Because the gauge-field ensembles omit electromagnetism, we need to subtract electromagnetic effects from the experimentally
measured masses, which means introducing a specific scheme to do so.
We identify $M_{\pi^0}^\text{QCD}=M_{\pi^0}^\text{expt}$ and adjust $m_l$ accordingly.
Values of the charged and neutral kaon masses in pure QCD, are obtained by subtracting the electromagnetic effects via the
quantities $(M^2_{K^0})^\gamma$ and $\epsilon'$.
Here $(M^2_{K^0})^\gamma$ is the electromagnetic contribution to the squared mass of the neutral kaon.
The quantity~$\epsilon'$ parametrizes higher-order corrections to Dashen's theorem:
\begin{equation}
    \epsilon' \equiv \frac{(M^2_{K^\pm}-M^2_{K^0})^\gamma - (M^2_{\pi^\pm}-M^2_{\pi^0})^\textrm{expt}}
        {(M^2_{\pi^\pm}-M^2_{\pi^0})^{\textrm{expt}}}.
    \label{eq:epsp-def}
\end{equation}

In this paper, we use the most recent values from the MILC Collaboration \cite{Basak:2018yzz}: 
\begin{align}
    \epsilon^\prime &= 0.74(1)_{\rm stat}({}^{+\phantom{1}8}_{-11})_{\rm syst},
    \label{eq:eps-all-errors} \\
    (M^2_{K^0})^\gamma &= 44(3)_{\rm stat}(25)_{\rm syst}~\MeV^2 .
    \label{eq:K0-result}
\end{align}
Our scheme is the one introduced for $u$ and $d$ quarks in \rcites{Borsanyi:2013lga,Fodor:2016bgu}.
It defines the isospin limit in the presence of electromagnetism to be the point at which the masses of both the $u\bar u$ and $d\bar d$
pseudoscalar mesons (neglected quark-line-disconnected contributions) are equal to $M_{\pi^0}^\text{QCD}$.
The scheme is extended naturally to the $s$ quark using the fact that mass renormalization for staggered quarks is
multiplicative~\cite{Basak:2018yzz}.
Numerically, the scheme dependence predominantly affects $\left(M_{K^0}^2\right)^\gamma$ and has relatively little influence on the
value of $\epsilon'$.

Using \eqs{eps-all-errors}{K0-result}, $m_s$ is tuned to obtain
\begin{equation}
    \left(M_{K^+}^2+M_{K^0}^2-M_{\pi^0}^2\right)^\text{QCD} = \left(M_{K^+}^2+M_{K^0}^2-M_{\pi^0}^2\right)^\text{expt} -
        2\left(M_{K^0}^2\right)^\gamma - (1+\epsilon')\left(M_{\pi^+}^2-M_{\pi^0}^2\right)^\text{expt} .
    \label{eq:EM-kaon}
\end{equation}

As in \rcite{Bazavov:2017fBD}, we tune $m_c$ and $m_b$ with the phenomenological
formula~\cite{Rosner:1992qw,Goity:2007fu,Davies:2010ip}
\begin{equation}
     M^\text{expt}_{H_x} = M^\text{QCD}_{H_x} + A e_x e_h + B e_x^2,
     \label{eq:EM-model}
\end{equation}
where $A=4.44$~MeV, $B=2.4$~MeV~\cite{Bazavov:2017fBD} and $e_x$ and $e_h$ are charges of the valence light quark and heavy
antiquark, respectively.%
\footnote{In \eq{EM-model}, our scheme is defined by the dropping of any term proportional to $e_h^2 m_h$, which 
could arise from electromagnetic mass renormalization of the heavy quark.
However, our simple model also omits some physical effects, such as a term proportional to $e^2/m_h$,
which would come from electromagnetic corrections to the quark-gluon vertex.} %
Using these quantities, the quark charges, and the experimental meson masses $M^\text{expt}_{D_s}=1968.27(10)$~MeV and
$M^\text{expt}_{B_s}=5366.82(22)$~MeV~\cite{Olive:2016xmw}, we compute the pure QCD masses $M^\text{QCD}_{D_s}=1967.01$~MeV and
$M^\text{QCD}_{B_s}=5367.04$~MeV.
This choice for defining $M^\text{QCD}_{H_x}$ amounts to a specific QED renormalization scheme for the heavy-quark mass.
Another choice, for example, would be to subtract the leading QED contribution to the self-energy of the heavy quark, which is
proportional to~$e_h^2$.
Finally, we set $(aM_{H_s}/af_{p4s})^\text{sim}=M^\text{QCD}_{H_s}/f_{p4s}$ to find the physical $am_c$ and $am_b$ on each ensemble.

In the one-loop \chpt\ result, \eq{chiral-form:2p1}, finite-volume effects enter through the function $K_1$.
Because the numerical evaluation of those effects is time-consuming, our base fit, as well as various alternative fits that we
employ to estimate or check statistical and systematic errors, use the infinite-volume version of $K_1$.
The finite-volume correction is determined only in a single fit at the end of the analysis.
Cross terms between finite-volume and other systematic errors are missed with this approach, but they are negligible.

We use a constrained fitting procedure \cite{Lepage:2001ym} with priors set as follows.
For the main objectives of the analysis, we choose extremely wide priors: $0\pm 6$~GeV for both $m_{p4s,\MSbar}(2~\GeV)$ and
$\Lambdabar_\MRS$, and $(0\pm1)\LamHQET^2$ for $\mu_\pi^2$.
Several other parameters are set from external considerations.
As discussed in Sec.~\ref{sec:SChPT}, the value of $\mu_G^2(m_b)$ should be close to the $B^*$-$B$ hyperfine splitting; following
\rcite{Gambino:2013rza}, we set the prior distribution of $\mu_G^2(m_b)$ to $(0.35\pm0.07)~\GeV^2$.
For the LECs that appear at LO in HMrA\schpt\ and are common for both decay constants and meson masses, we use the same prior
constraints as in our work on decay constants~\cite{Bazavov:2017fBD}:
\begin{subequations}
    \label{eq:external:priors}
    \begin{align}
        g_\pi &\sim 0.53 \pm 0.08, \label{eq:priors:gpi}\\
        \frac{1}{f^2} &\sim \frac{1}{2}\left(\frac{1}{f_\pi^2}+\frac{1}{f_K^2}\right)\pm
            \left(\frac{1}{f_\pi^2}-\frac{1}{f_K^2}\right), \label{eq:priors:logcoeff}\\
        \delta'_V/\bar{\Delta} &\sim -0.88\pm0.09, \\
        \delta'_A/\bar{\Delta} &\sim +0.46\pm0.23,
    \end{align}
\end{subequations}
where $a^2\bar{\Delta}$ is related to the differences in squared pion masses, as discussed in Ref.~\cite{Bazavov:2017fBD}.
For the LECs $\lambda_1$ and $\lambda'_1$, we use wide priors of $(0\pm2)~\GeV^{-1}$, which are 10 times larger than
what can be extracted from the flavor splittings of $B$ or $D$ mesons, namely
$\lambda_1\approx0.2~\GeV^{-1}$ (see, for example, Ref.~\cite{Bazavov:2011aa}).
Similarly, for the dimensionless LECs $\lambda_{a^2}$ and $\lambda'_{a^2}$, we use priors of $0\pm10$,
which are much wider than the expected size of order~1.

For the overall normalization of $R_n$, which is denoted by $N_m$ in Ref.~\cite{Pineda:2001zq} and $N$ in
Refs.~\cite{Komijani:2017vep,Beneke:1994rs}, we use
\begin{equation}
    R_0 = 0.535\pm0.010
    \label{eq:renormalon:N}
\end{equation}
for a theory with three massless active quarks~\cite{Komijani:2017vep}.
In the fits reported in this section, we use Eq.~(\ref{eq:renormalon:N}) to provide a prior for~$R_0$.

Finally, the remaining parameters, which are dimensionless, are given the prior $0\pm1$.

The calculation of the MRS mass relies on having a precise estimate for the strong coupling.
In this paper, we use 
\begin{equation}
    \alpha_\MSbar(5~\GeV; n_f=4) = 0.2128(25),
    \label{eq:alpha5GeV}
\end{equation}
which has been obtained by HPQCD Collaboration~\cite{Chakraborty:2014aca,*Chakraborty:2017aca} for four active flavors.
This value corresponds to $\alpha_\MSbar(m_Z; n_f=5)=0.11822(74)$, whereas the PDG quotes $\alpha_\MSbar(m_Z; n_f=5)=0.1181(11)$
with a somewhat larger uncertainty.
The advantage of Eq.~(\ref{eq:alpha5GeV}) is that it has been determined on a subset of the same ensembles used here, so it is
consistent to use it with our lattice-QCD data.
We use the mean value in our base fit, and we introduce an uncertainty associated with $\alpha_\MSbar$ by varying its value by
$1\sigma$.
To run the coupling constant to the scale $\mu$, we use the QCD beta function at five-loop order accuracy~\cite{Baikov:2016tgj} and
integrate the differential equation numerically.
In Sec.~\ref{sec:Results}, we comment on how the results would change using the PDG's estimate of the uncertainty in~$\alpha_\MSbar$.

In general, our data for the meson masses are more precise than the data for scale-setting quantities.
We incorporate the latter uncertainties as follows.
Let us use $af_{p4s}$ and $am_{p4s}$ to denote the $p4s$ quantities computed from light mesons at each lattice spacing and
$\Sigma_{p4s}$ to denote their covariance matrix.
We introduce two fit parameters at each lattice spacing, $af_{p4s,\text{opt}}$ and $am_{p4s,\text{opt}}$, to represent optimized
values for the $p4s$ quantities under the influence of the heavy-light data.
We then employ the so-called penalty trick~\cite{Ball:2009qv} to take into account the uncertainties in $af_{p4s}$ and $am_{p4s}$.
Thus, we add
\begin{equation}
    \delta \chi^2 = \sum
    \begin{bmatrix} af_{p4s}-af_{p4s,\text{opt}}\quad{} & am_{p4s}-am_{p4s,\text{opt}} \end{bmatrix}
        \left(\Sigma_{p4s}\right)^{-1}
    \begin{bmatrix} af_{p4s}-af_{p4s,\text{opt}}     \\   am_{p4s}-am_{p4s,\text{opt}} \end{bmatrix}
\end{equation}
to our $\chi^2$ function, where the sum is over all lattice spacings.
Because data at 5 different lattice spacings enter the base fit, 10 additional parameters are required.
The optimized values for the scale setting quantities are then obtained simultaneously in the EFT fit.
Given the size of errors in our data, the bias discussed in \rcite{Ball:2009qv} is negligible.

Altogether we have 384 lattice data points and 77 parameters in our base fit: 67 parameters in the EFT fit function
and 10 parameters for optimized values of scale-setting quantities.
The fit returns a correlated $\chi^2_{\text{data}}/\text{dof}=320/307$, giving a $p$~value of $p=0.3$.
\begin{figure}
    \includegraphics[width=0.48\textwidth]{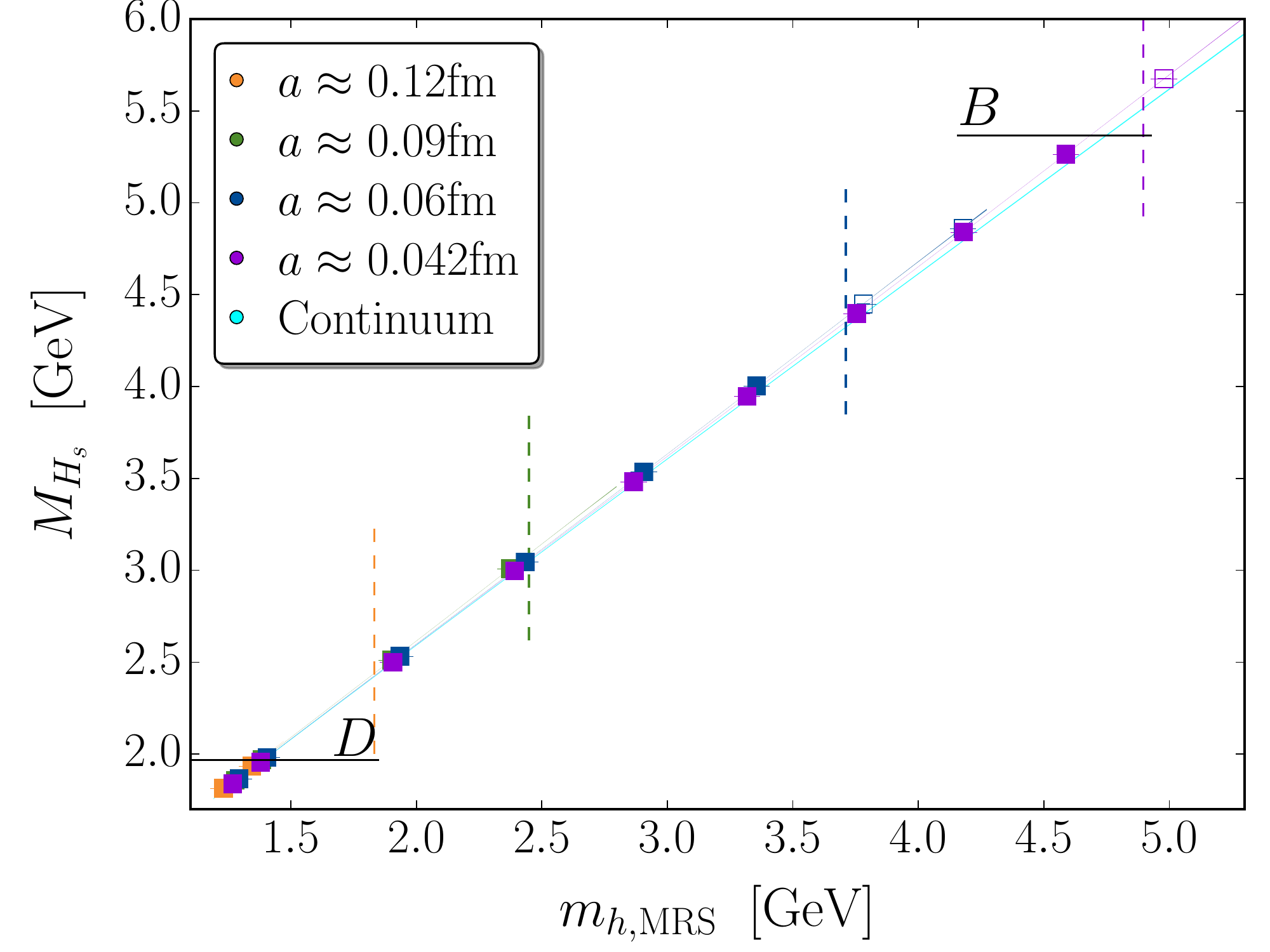} \hfill
    \includegraphics[width=0.48\textwidth]{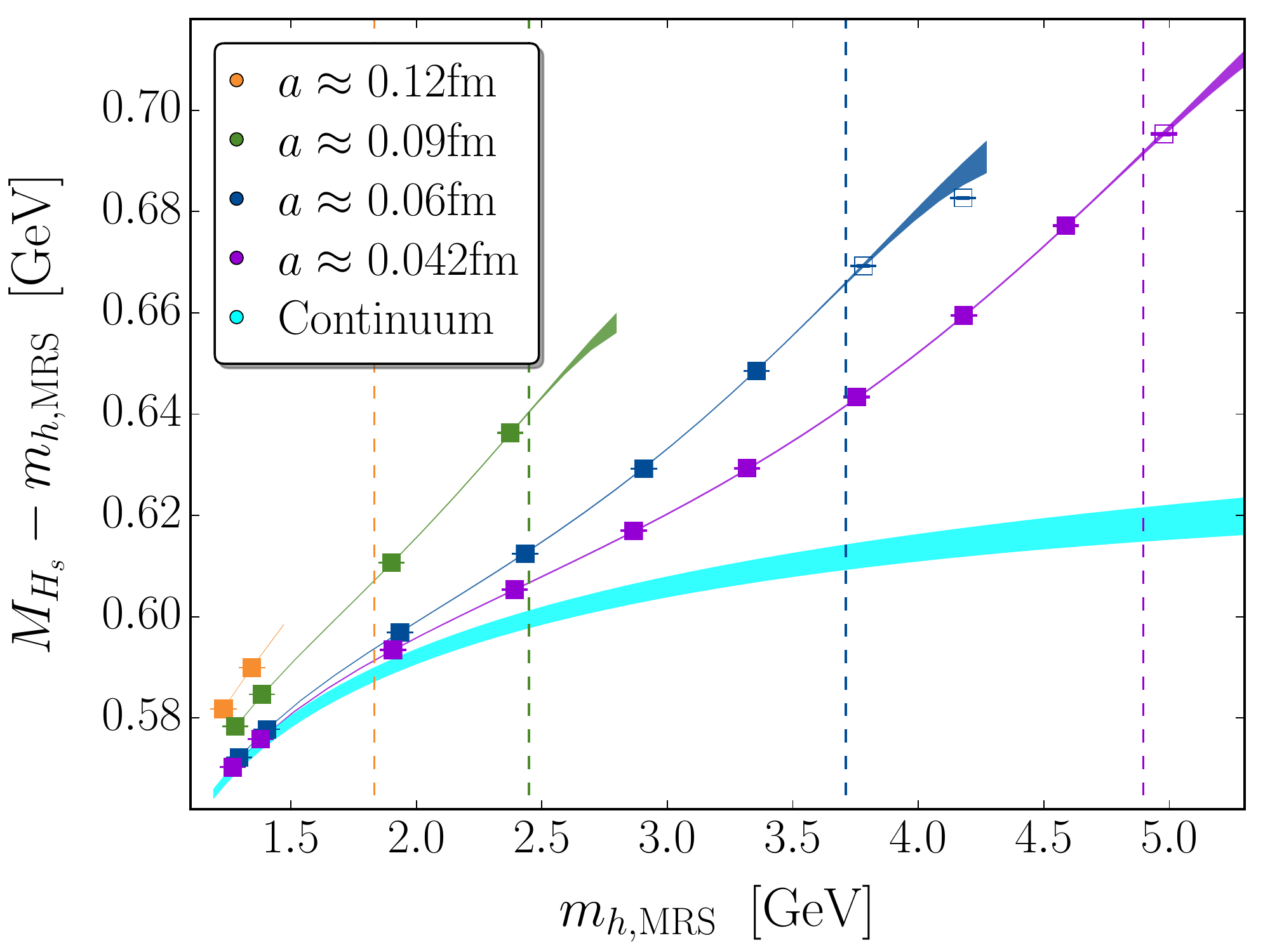}
    \caption{A snapshot of the base fit (to data for all light-quark masses) and the lattice data for heavy-strange meson masses.
	Only ensembles with physical light sea mass are shown, thereby leaving out the finest lattice
        spacing, $a\approx0.03$~fm.
        Left:~heavy-strange meson mass vs.\ heavy-quark MRS mass.
        Right:~difference of the heavy-strange meson mass and the heavy-quark MRS mass vs.\ heavy-quark MRS mass.
        The dashed vertical lines indicate the cut $am_h = 0.9$ for each lattice spacing.
        Data points with open symbols to the right of the dashed vertical lines are omitted from the fit.
        Here $m_{h,\MRS}$ is the continuum limit of the MRS mass of the heavy quark~$h$.
        The error bar for $m_{h,\MRS}$ is suppressed for clarity.}
    \label{fig:mass-vs-mMRS}
\end{figure}
\Figrefs{mass-vs-mMRS}{mass-vs-mMRSinv} illustrate the base fit at the four (five) lattice spacings for the physical
mass ($0.2 m'_s$) ensembles and in the continuum limit.
\begin{figure}
    \includegraphics[width=0.48\textwidth]{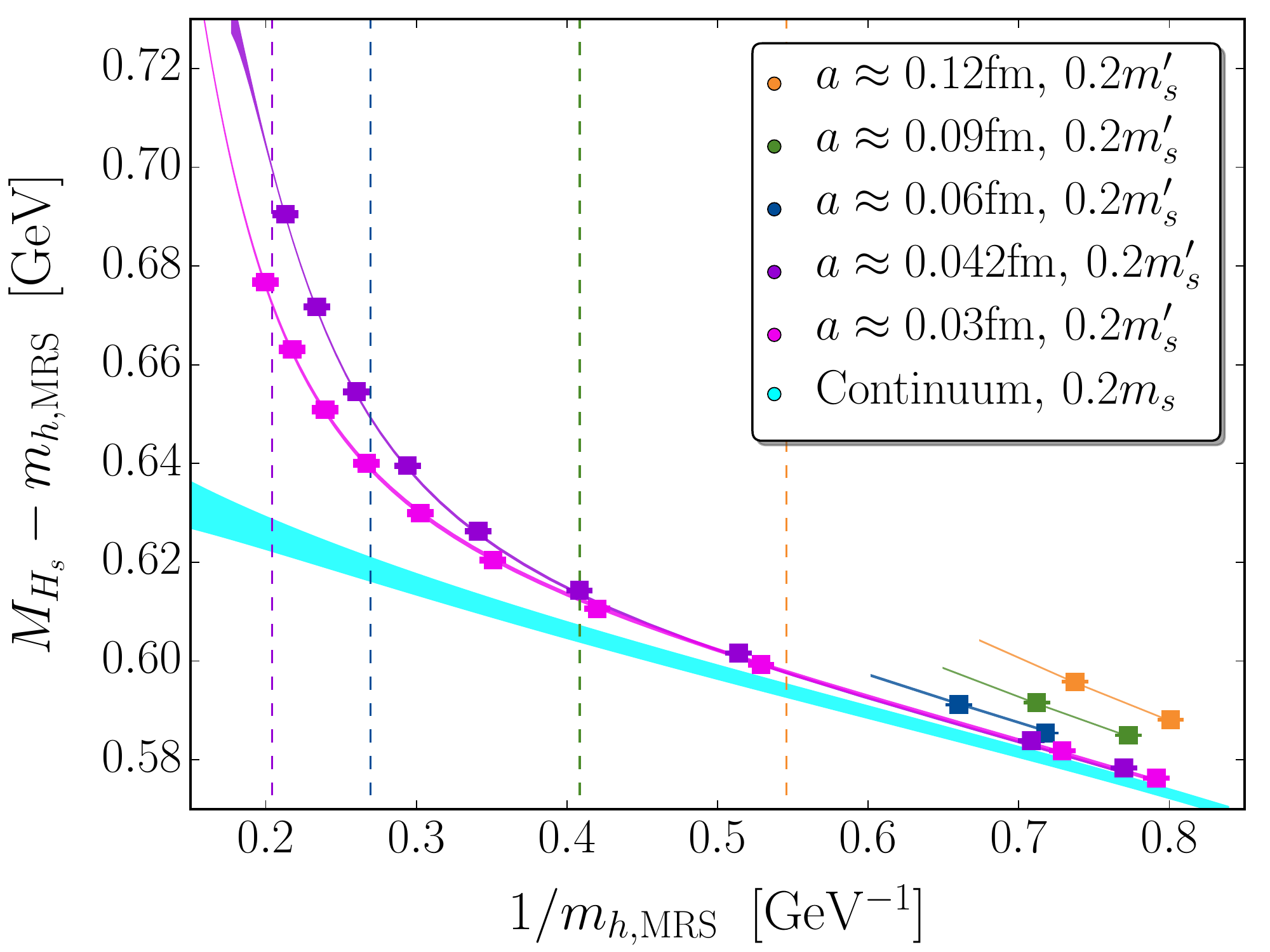} \hfill
    \includegraphics[width=0.48\textwidth]{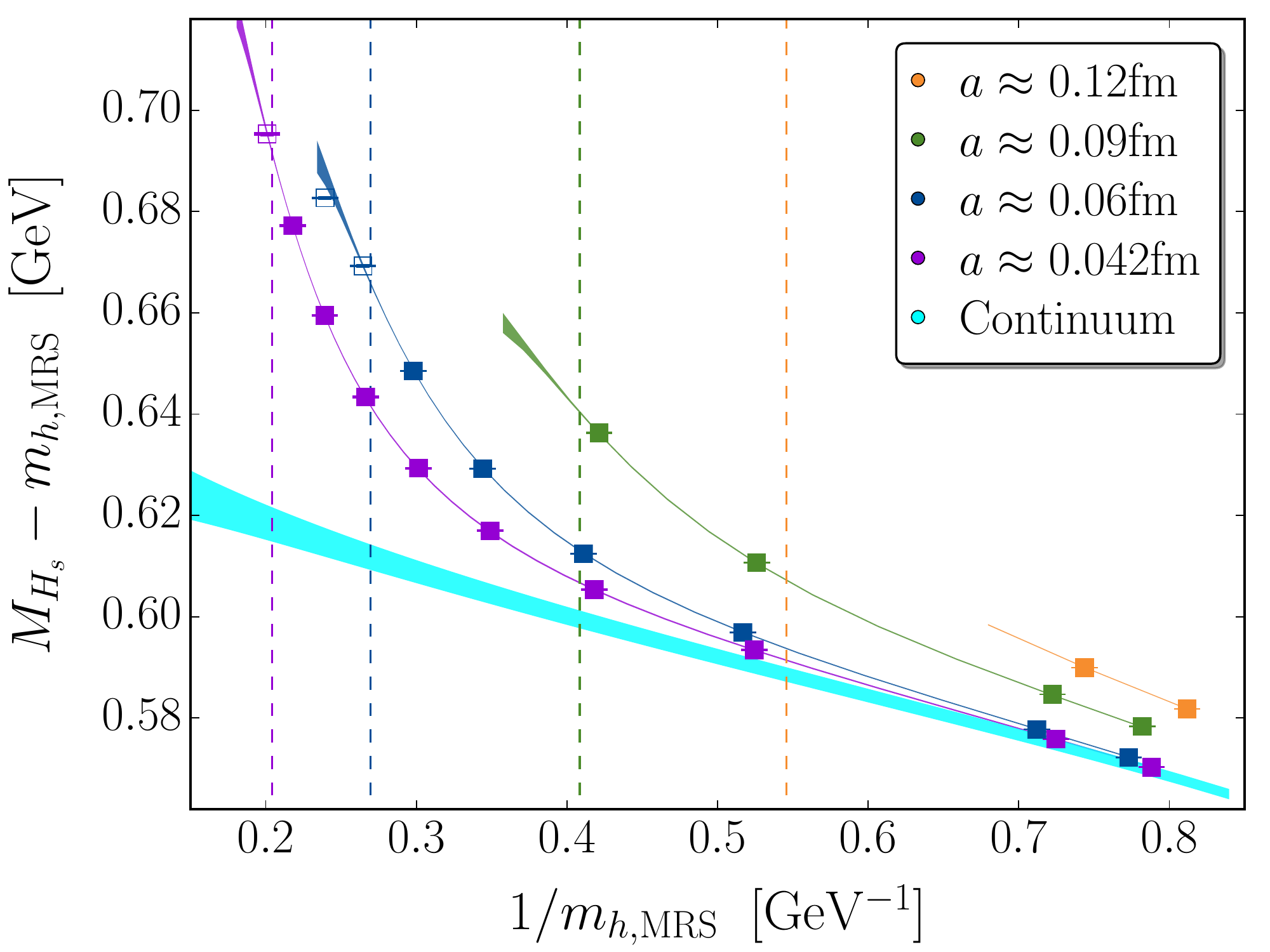}
    \caption{Similar to \figref{mass-vs-mMRS}, but the horizontal axes are the inverse of the heavy-quark MRS mass,
        and the left panel shows ensembles with light sea mass equal to $0.2 m'_s$,
        while the right panel shows physical-mass ensembles.
        Data points with open symbols to the left of the dashed vertical lines are omitted from the fit.}
    \label{fig:mass-vs-mMRSinv}
\end{figure}
The valence light mass $m_x$ is tuned to $m_s$: the graphs illustrate a snapshot for heavy-strange meson masses.
We plot the heavy-strange meson mass or the difference of the meson mass and the $\bar{h}$-antiquark MRS
mass versus the continuum limit of the $h$-quark MRS mass (in \figref{mass-vs-mMRS}) or  its reciprocal (in
\figref{mass-vs-mMRSinv}).
Data points with open symbols to the right (left) of the dashed vertical line of the corresponding color in \figref{mass-vs-mMRS}
(\figref{mass-vs-mMRSinv}) are omitted from the fit because they have $am_h>0.9$.
In the continuum extrapolation the masses of sea quarks are set to the physical (correctly tuned) quark masses $m_l$, $m_s$ and
$m_c$, while at nonzero lattice spacing the masses of the sea quarks take their simulation values.

The width of the fit lines in \figrefs{mass-vs-mMRS}{mass-vs-mMRSinv} show the statistical error coming from the fit, which is only
part of the total statistical error, since it does not include the statistical errors in the inputs of the light quark masses and
the lattice scale.
Furthermore, the statistical error reported by the fit is sensitive to numerical errors in computing the fit parameters' covariance
matrix.
For a robust determination of the total statistical error of each output quantity, we divide the full data set into 20 jackknife
resamples.
The complete calculation, including the determination of the inputs, is performed on each resample, and the error is computed as
usual from the variations over the resamples.
For convenience, we keep the covariance matrix fixed to that of the full data set, rather than recomputing it for each resample.

The physically interesting quantities $m_{p4s,\MSbar}(2~\GeV)$, $\Lambdabar_\MRS$, $\mu_\pi^2$, and $\mu_G^2(m_b)$ are now
determined directly from the fit to the lattice data.
Moreover, the fit function evaluated at zero lattice spacing and physical sea-quark masses yields the meson masses as a function of
the valence heavy and light quark masses; see, \eg Figs.~\ref{fig:mass-vs-mMRS} and~\ref{fig:mass-vs-mMRSinv}.

\begin{figure}
  \includegraphics[width=1\textwidth]{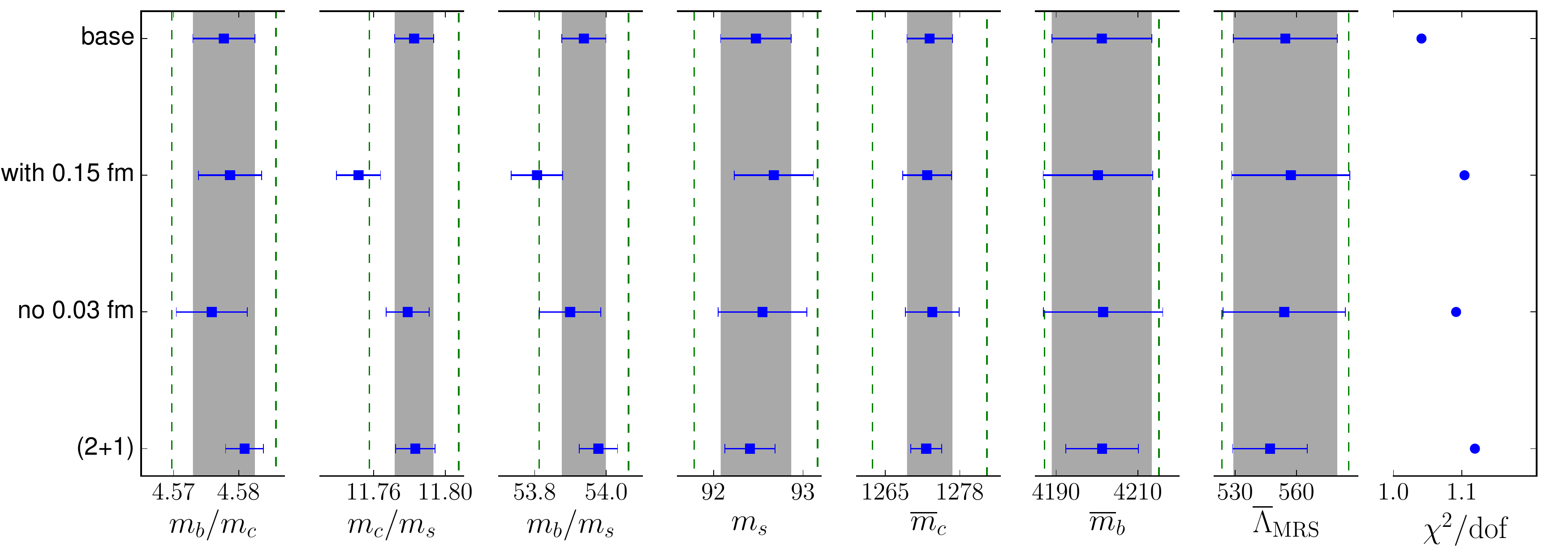}
  \caption{Stability plot showing the sensitivity under variations in the data set and the form of the fit function,
    as described in the text.
    Here $m_s=m_{s,\MSbar}(2~\GeV)$, $\mcbar = m_{c,\MSbar}(m_{c,\MSbar})$, and $\mbbar = m_{b,\MSbar}(m_{b,\MSbar})$.
    The error bars show only the statistical errors, the gray error bands correspond to the statistical error of the base fit,
    and the dashed green lines correspond to total errors.}
\label{fig:stability}
\end{figure}

\Figref{stability} shows the stability of our final results for $\MSbar$ quark mass ratios $m_b/m_c$, $m_c/m_s$ and $m_b/m_c$;
for masses of strange, charm and bottom quarks; and for the HQET matrix element $\Lambdabar_\MRS$.
We test the systematic error in the continuum extrapolation by repeating the fit after either adding in the coarsest
($a\approx0.15$~fm) ensembles or omitting the finest ($a\approx0.03$~fm) ensemble.
These changes are shown in \figref{stability} and seen to have no significant effect, so we consider these tests to be
cross-checks.
The meson-mass data in our base fit are obtained from the (3+2)-state fits to two-point correlators.
To investigate the error arising from excited state contamination, we repeat the EFT fit with meson-mass data from the (2+1)-state
fits to two-point correlators.
As seen in \figref{stability}, the effects from this change are small too.
Because we have no other handle on systematic errors due to excited states, we take the difference between the results from the two
types of correlator fits as an estimate of this uncertainty.

We now turn to effects from truncating perturbative QCD in the relation between quark-mass definitions and the beta function.
As explained with Eq.~(\ref{eq:rabbit}), the MRS mass connects the $\MSbar$ mass of the $h$ quark to the heavy-light-meson
mass~$H_x$.
By design, the fit yields $m_{p4s,\MSbar}(2~\GeV)$, and we use the continuum limit of $am_h/am_{p4s}$ to convert to
$m_{h,\MSbar}(2~\GeV)$.
We then use \eqs{mMSbar2mRGI}{mMSbar2mRGI_c} to calculate $\mbar_h$ and \eq{mSI2mMRS} to calculate $m_{h,\MRS}$.
The beta function and quark-mass anomalous dimension are known at five loops~\cite{Baikov:2014qja,Baikov:2016tgj}, and the pole mass
at four loops~\cite{Marquard:2015qpa,Marquard:2016dcn}.

\begin{figure}
    \includegraphics[width=1\textwidth]{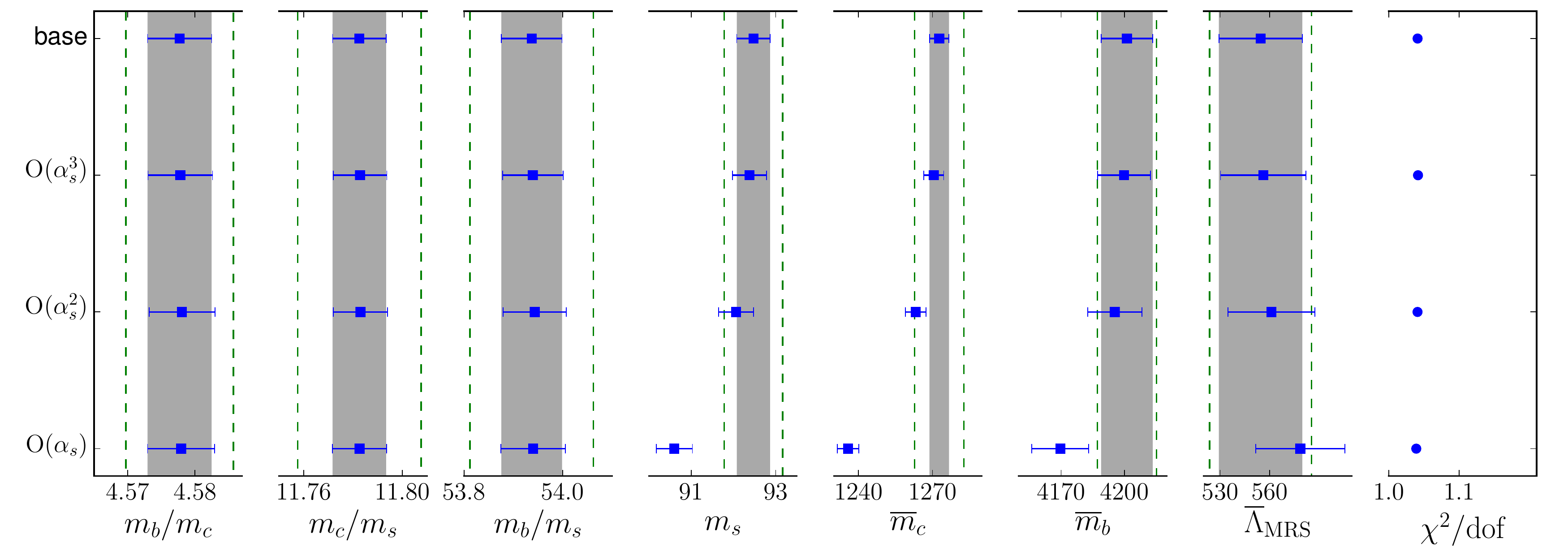}
    \caption{Stability plot showing the sensitivity to truncation error in perturbative-QCD relations that are used in our analysis.
        In the base fit, the perturbative series are accurate through order~$\alpha_s^4$.
        In the fits labeled by $\order(\alpha_s^n)$, we keep $n$ subleading orders.
        Here $m_s=m_{s,\MSbar}(2~\GeV)$, $\mcbar = m_{c,\MSbar}(m_{c,\MSbar})$, and $\mbbar = m_{b,\MSbar}(m_{b,\MSbar})$.
        The error bars show only the statistical errors, the gray error bands correspond to the statistical error of the base fit,
        and the dashed green lines correspond to total errors.}
   \label{fig:stability_loop}
\end{figure}

To monitor the errors from truncating perturbative QCD, we rerun the analysis with fewer orders in \eqs{mMSbar2mRGI_c}{mSI2mMRS} and
in the beta function without, however, changing $C_\text{cm}$, the Wilson coefficient for $\mu^2_G$,
\eq{C_cm}.
\Figref{stability_loop} shows the stability of our results for $\MSbar$ quark mass ratios $m_b/m_c$, $m_c/m_s$ and $m_b/m_c$; for
masses of strange, charm and bottom quarks; and for the HQET matrix element~$\Lambdabar_\MRS$, as the order of perturbation theory
is increased.
In Fig.~\ref{fig:stability_loop}, we denote by $\order(\alpha_s^n)$ a fit that includes $n$ orders beyond the leading terms in
Eqs.~(\ref{eq:mMSbar2mRGI_c}), (\ref{eq:mSI2mMRS}), and the beta function.
The quark mass ratios are not at all sensitive to the truncations in the perturbative-QCD relations; essentially, these ratios are
the continuum limit of the corresponding bare masses.
For the quark masses and the HQET matrix elements, one finds good convergence, within the statistical errors, as the order of
$\alpha_s$ in the perturbative expressions is increased.
Based on this observation, we do not introduce any additional systematic error associated with truncation in perturbative-QCD
results.
Note that the truncation effects are negligible because the renormalon-subtracted perturbative coefficients in the MRS mass are
all very small.
If one employs the RS mass~\cite{Pineda:2001zq}, for instance, the truncation error for the bottom quark mass would be
about 10 to 20~MeV, depending on the details.

\begin{figure}
    \includegraphics[width=1\textwidth]{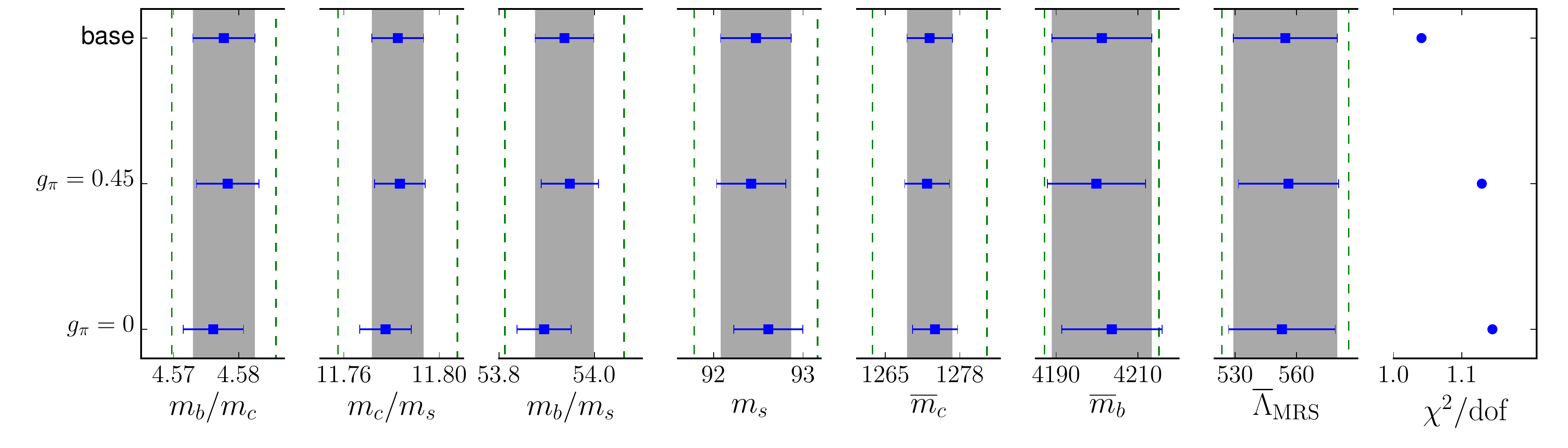}
    \caption{Stability plot showing the sensitivity to different choices for $g_\pi$.
    The error bars show only the statistical errors, the gray error bands correspond to the statistical error of the base fit 
    and the dashed green lines correspond to total errors.}
    \label{fig:stability_gpi}
\end{figure}

Our data prefer an overall coefficient of the one-loop \aschpt\ contribution, $\delta M_{H_x}$, namely $g_\pi^2/f^2$, well below the
prior width of the product of $g_\pi^2$ and $1/f^2$ in \eqs{priors:gpi}{priors:logcoeff}.
In our base fit, the posterior for this product is $g_\pi^2/f^2 = 3.8~\GeV^{-2}$ (for $D$ systems) while the prior value is
$(14\pm7)~\GeV^{-2}$.
To investigate the effects of treating $g_\pi^2$ and $1/f^2$ as free parameters, we consider two alternative fits.
First, we fix $g_\pi=0.45$, one sigma below its nominal value, and $1/f^2=1/f_K^2$.
Second, we fix $g_\pi=0$, which is equivalent to fitting to a polynomial in the quark masses.
In \figref{stability_gpi}, we label the first of these ``$g_\pi=0.45$'' and the second ``$g_\pi=0$''.
As one can see, the quark-mass ratios, quark masses themselves, and the HQET matrix element $\Lambdabar_\MRS$ do not change
significantly under these variations.
Consequently, we do not introduce any additional systematic error associated with our treatment of $g_\pi$ and~$1/f^2$.

Our full error budget for quark-mass ratios, quark masses, and the HQET matrix element $\Lambdabar_\MRS$ is given in
Table~\ref{tab:results_masses_details}.
The row labeled ``statistics and EFT fit'' lists the uncertainty reported by the Bayesian fit, which incorporates associated
systematic effects of extrapolation.
There are further systematic effects not captured in the EFT fit.
The excited-state contamination in two-point correlator fits is explained above.
Our method of estimating the systematic error associated with the tuned quark masses and scale setting quantities is similar to that
in \rcite{Bazavov:2017fBD}.
We correct for (exponentially small) finite-volume effects using the finite-volume version of the NLO \chpt\ for the heavy-light
mesons, and using NLO or NNLO \chpt\ for the light-quark and scale-setting inputs following \rcite{Bazavov:2014wgs}.
Residual finite-volume effects from higher orders in \chpt\ are estimated, as in \rcite{Bazavov:2017fBD}, as 0.3 times the
calculated finite-volume correction.
The nonequilibration of topological charge in our finest ensembles causes small finite-volume effects that are not exponentially
suppressed~\cite{Bernard:2017npd}.
Although this error is negligible for masses of heavy-light mesons, even with our high statistics, we include the shift expected
from Ref.~\cite{Bernard:2017npd} as a systematic error.
Note that, despite the fact that we take the full topological shift as the associated error, these errors all round to zero at
the precision shown in Table~\ref{tab:results_masses_details}.
The uncertainties stemming from the omission of electromagnetism are discussed in detail below.
Finally, our results have uncertainties from the parametric inputs $\alpha_s$, given in Eq.~(\ref{eq:alpha5GeV}), and
$\fpiPDG=130.50(13)$~MeV~\cite{Rosner:2015wva}.


\begin{table}
\newcommand{\h}{\phantom{x}}
\caption{Error budget for strange-, charm- and bottom-quark masses, their ratios, and the HQET matrix element~$\Lambdabar_\MRS$.
  See the text for the description.}
\label{tab:results_masses_details}
\begin{tabular*}{\textwidth}{@{\extracolsep{\fill}}lccccccc}
\hline\hline
Error (\%) &  $m_b/m_c$  &  $m_c/m_s$  &  $m_b/m_s$  &  $m_{s,\MSbar}(2~\GeV)$  &  $\mbar_c$  &  $\mbar_b$  &  $\Lambdabar_\MRS$  \\
\hline
Statistics and EFT fit	  & $0.10$ & $0.09$ & $0.11$ & $0.43$ & $0.31$ & $0.29$ & $4.6$ \\
Two-point correlator fits & $0.07$ & $0.01$ & $0.08$ & $0.07$ & $0.05$ & $0.00$ & $1.3$ \\
Scale setting and tuning  & $0.02$ & $0.14$ & $0.16$ & $0.18$ & $0.03$ & $0.02$ & $0.2$ \\
Finite-volume corrections & $0.00$ & $0.02$ & $0.01$ & $0.01$ & $0.01$ & $0.00$ & $0.0$ \\
Topological charge distribution & $0.01$ & $0.00$ & $0.00$ & $0.01$ & $0.01$ & $0.01$ & $0.2$ \\
Electromagnetic corrections     & $0.12$ & $0.11$ & $0.01$ & $0.01$ & $0.08$ & $0.00$ & $0.0$ \\
$\alpha_s$                      & $0.01$ & $0.00$ & $0.01$ & $0.56$ & $0.75$ & $0.18$ & $2.9$ \\
$\fpiPDG$                       & $0.03$ & $0.07$ & $0.10$ & $0.12$ & $0.04$ & $0.02$ & $0.2$ \\
\hline\hline
\end{tabular*}

\end{table}

\begin{table}
    \caption{Error contributions from electromagnetic effects to strange-, charm- and bottom-quark
        masses, their ratios, and the HQET matrix element~$\Lambdabar_\MRS$.
        The sources of uncertainty are described in the text.}
    \label{tab:results_masses_EM}
    \begin{tabular}{lccccccc}
        \hline\hline
        Error (\%) & $m_b/m_c$ & $m_c/m_s$ & $m_b/m_s$ & $m_{s,\MSbar}(2~\GeV)$ & $\mbar_c$ & $\mbar_b$ & $\Lambdabar_\MRS$\\
        \hline
	\EMone   & $+0.00$ & $-0.01$ & $-0.01$ & $+0.00$ & $-0.00$ & $-0.00$ & $+0.0$ \\
	\EMtwo   & $-0.00$ & $+0.00$ & $+0.00$ & $-0.01$ & $-0.00$ & $-0.00$ & $+0.0$ \\
	\EMthree & $-0.12$ & $+0.11$ & $-0.00$ & $+0.00$ & $+0.08$ & $-0.00$ & $-0.0$ \\
        \hline\hline
    \end{tabular}
\end{table}


Table~\ref{tab:results_masses_EM} shows a breakdown of the uncertainties from matching a pure-QCD calculation such as this to
QCD+QED.
Briefly, ``\EMone'' is the uncertainty in connecting the $K^+$-$K^0$ splitting to that of $\pi^+$-$\pi^0$, stemming from
$\epsilon'$, and ``\EMtwo'' refers to the uncertainty in the electromagnetic contribution to the neutral kaon mass,
$\left(M_{K^0}^2\right)^\gamma$.
These two effects are negligible compared with the other sources of uncertainty.
In this work, we choose a specific scheme~\cite{Borsanyi:2013lga,Basak:2018yzz} for the electromagnetic contribution to the neutral kaon
masses; other works, for example the FLAG report~\cite{Aoki:2016frl}, choose other schemes.
Changing $\left(M_{K^0}^2\right)^\gamma$ from $+44~\MeV^2$ to $+461~\MeV^2$ reduces $m_s$ by
0.17\% and, consequently, increases $m_c/m_s$ and $m_b/m_s$ by 0.17\%.
When using these ratios and the strange-quark mass in a setting that ignores the subtleties of the QED scheme, one may wish to
incorporate an additional uncertainty of $\pm0.17\%$.

Another uncertainty comes from the estimates of the electromagnetic correction to the heavy-light meson mass, described above with
Eq.~(\ref{eq:EM-model}).
It is denoted ``\EMthree'' in Table~\ref{tab:results_masses_EM}.
For the associated error, we take the difference between results obtained with and without the electromagnetic shift.
Results for heavy-quark masses depend on the chosen QED quark-mass scheme.
As discussed above, we do not subtract any part of the QED self-energy.
[Equation~(\ref{eq:EM-model}) contains no term proportional to $e_h^2$.] %
When using other schemes, one should convert our results accordingly: a shift of 1~MeV in the QCD part of the $D_s$ ($B_s$) mass
leads to a 0.7~MeV (0.8~MeV) shift in $\mbar_c$ ($\mbar_b$).
The scheme dependence on the meson masses may be estimated as $\pm \alpha m_he_h^2\approx\pm4.2$~MeV (average for $D_s$ and $B_s$).
When using the heavy-quark masses in a setting that ignores the subtleties of the QED scheme, one may consequently wish to
incorporate an additional uncertainty of $\pm3.1$~MeV on $\mbar_c$ and $\pm3.5$~MeV on~$\mbar_b$.

\section{Results}
\label{sec:Results}

In this section, we collect the results that stem from the EFT fits described in the previous two sections.
These fall into four categories: quark masses themselves and their ratios, HQET matrix elements, flavor splittings in the $D$ and 
$B$ systems, and LECs of \hmchpt.
We emphasize again that our final results for quark masses depend on our prescription for calculating QCD-only meson masses; cf.\
the discussions around Eq.~(\ref{eq:EM-model}) and about \tabref{results_masses_EM}.

\subsection{Quark masses}
\label{subsec:masses}

As discussed in Sec.~\ref{sec:HQET-SChPT_fit}, the main physical fit parameters correspond to the terms in Eq.~(\ref{eq:mQ_2_MH}).
For the masses, the fit yields
\begin{equation}
    m_{p4s,\MSbar}(2~\GeV) = 36.99 (16)_\text{stat} (07)_\text{syst} (21)_{\alpha_s} (04)_{\fpiPDG}~\MeV
    \label{eq:mp4s_MS}
\end{equation}
with four active flavors, from which it follows immediately
\begin{equation}
    m_{s,\MSbar}(2~\GeV) = 92.47 (39)_\text{stat} (18)_\text{syst} (52)_{\alpha_s} (11)_{\fpiPDG}~\MeV .
    \label{eq:ms_MS}
\end{equation}
Having determined the strange quark mass, we use the quark mass ratios $m_s/m_l$ and $m_u/m_d$,
and their correlations, to obtain the light quark masses
\begin{align}
    m_{l,\MSbar}(2~\GeV) &= 3.402 (15)_\text{stat} (05)_\text{syst} (19)_{\alpha_s} (04)_{\fpiPDG}~\MeV ,
    \label{eq:ml_MS} \\
    m_{u,\MSbar}(2~\GeV) &= 2.130 (18)_\text{stat} (35)_\text{syst} (12)_{\alpha_s} (03)_{\fpiPDG}~\MeV ,
    \label{eq:mu_MS} \\
    m_{d,\MSbar}(2~\GeV) &= 4.675 (30)_\text{stat} (39)_\text{syst} (26)_{\alpha_s} (06)_{\fpiPDG}~\MeV ,
    \label{eq:md_MS}
\end{align}
where $m_l$ is again the average of the up- and down-quark masses.
To obtain these results, we take the large side of the asymmetric uncertainties reported in 
\rcite{Bazavov:2017fBD}, namely, $m_s/m_l=27.178(47)_\text{stat}(70)_\text{syst}(1)_{\fpiPDG}$ and
$m_u/m_d=0.4556(55)_\text{stat}(114)_\text{syst}(0)_{\fpiPDG}$.

Evaluating the fit function at the quark masses yielding the $D_s$ and $B_s$ mesons yields the mass ratios
\begin{align}
    m_c/m_s  &= 11.783 (11)_\text{stat} (21)_\text{syst} (00)_{\alpha_s} (08)_{\fpiPDG} ,
    \label{eq:mc_ms}\\
    m_b/m_s  &= 53.94   (6)_\text{stat} (10)_\text{syst}  (1)_{\alpha_s}  (5)_{\fpiPDG} ,
    \label{eq:mb_ms}\\
    m_b/m_c  &=  4.578  (5)_\text{stat}  (6)_\text{syst}  (0)_{\alpha_s}  (1)_{\fpiPDG}
    \label{eq:mb_mc}
\end{align}
where the third line is the ratio of the first two, taking correlations in the uncertainties into account.
In $m_c/m_s$ and $m_b/m_c$, the uncertainty stemming from $\alpha_s$ rounds to zero.
As elsewhere in this paper, these quark-mass ratios are given in our scheme for subtracting
electromagnetic contributions from the $K^0$, $D_s$, and $B_s$ meson masses.
With this proviso in mind, though, they hold for any mass-independent renormalization scheme of QCD.

For the charm- and bottom-quark masses we then obtain
\begin{align}
    m_{c,\MSbar}(2~\GeV) &= 1090  (5)_\text{stat} (2)_\text{syst} (6)_{\alpha_s} (1)_{\fpiPDG}~\MeV ,
    \label{eq:mc_MS-2GeV} \\
    m_{b,\MSbar}(2~\GeV) &= 4988 (17)_\text{stat} (1)_\text{syst}(29)_{\alpha_s} (1)_{\fpiPDG}~\MeV ,
    \label{eq:mb_MS-2GeV}
\end{align}
again for four active flavors.
The relative systematic error is larger for $m_{c,\MSbar}(2~\GeV)$ than for $m_{b,\MSbar}(2~\GeV)$, because much of it comes from
additive parts of the two-point correlator and electromagnetic uncertainties.
The largest uncertainty comes from the uncertainty in $\alpha_s$ in Eq.~(\ref{eq:alpha5GeV}), followed by the statistical error
(after propagation through the EFT fit).
As one can see from Fig.~\ref{fig:stability_loop} and Eq.~(\ref{eq:MRS-MSbar}), below, this uncertainty does not come from
order-by-order changes in perturbative QCD: the $\alpha_s$ uncertainty is parametric.

The uncertainty stemming from $\alpha_s$ becomes smaller at higher renormalization points.
For the charmed quark,
\begin{equation}
    m_{c,\MSbar}(3~\GeV)  = 983.7 (4.3)_\text{stat} (1.4)_\text{syst}  (3.3)_{\alpha_s} (0.5)_{\fpiPDG}~\MeV,
    \label{eq:mc_MS-3GeV}
\end{equation}
or, adding all errors in quadrature, $983.7(5.6)~\MeV$.
Running from one renormalization scale is carried out with Eq.~(\ref{eq:mMSbar2mRGI_c}) and numerical integration of the
differential equation for $\alpha_\MSbar$ with the five-loop beta function.
For comparison to the literature (cf.\ Sec.~\ref{sec:Outlook}), it is useful to have
$\mbar_h = m_{h,\MSbar}(m_{h,\MSbar})$; for charm and bottom
\begin{align}
    \mbar_c  &= 1273  (4)_\text{stat} (1)_\text{syst} (10)_{\alpha_s}  (0)_{\fpiPDG}~\MeV ,
    \label{eq:mc_SI}\\
    \mbar_b  &= 4201 (12)_\text{stat} (1)_\text{syst}  (8)_{\alpha_s}  (1)_{\fpiPDG}~\MeV ,
    \label{eq:mb_SI}
\end{align}
or, adding all errors in quadrature, $\mbar_c=1273(10)$~MeV and $\mbar_b=4201(14)$~MeV.

The quark masses given above are for four active flavors.
The mass of the bottom quark  with five active flavors can be calculated from~\cite{Liu:2015fxa}
\begin{align}
    m_b^{(n_l)}(\mu) &= \mbbar^{(n_f)} \left[1 + 0.2060 \left(\frac{\alpha_s^{(n_f)}(\mu)}{\pi}\right)^2
        + (1.8476 + 0.0247 n_l) \left(\frac{\alpha_s^{(n_f)}(\mu)}{\pi}\right)^3 
    \right. \nonumber \\ \hspace{3em} & \left.
        + (6.850 - 1.466 n_l+0.05616 n_l^2)\left(\frac{\alpha_s^{(n_f)}(\mu)}{\pi}\right)^4 + \cdots \right] ,
\end{align}
where $n_l=n_f-1$ and $\mu=\mbbar^{(n_f)}$.
Setting $n_f=5$, we obtain
\begin{equation}
    \mbar_b^{(n_f=5)} = 4195 (12)_\text{stat} (1)_\text{syst}  (8)_{\alpha_s}  (1)_{\fpiPDG}~\MeV ,
    \label{eq:mb_SI_nf5}
\end{equation}
or, adding all errors in quadrature, $\mbar_b=4195(14)$~MeV.
The five-flavor mass can be run from $\mbar_b^{(n_f=5)}$ to higher scales using the five-loop anomalous
dimension~\cite{Baikov:2014qja} and beta function~\cite{Baikov:2016tgj} with $n_f=5$.
For completeness, we run the $b$ mass to 10~GeV, finding
\begin{equation}
    m_{b,\MSbar}(10~\GeV; n_f=5) = 3665 (11)_\text{stat} (1)_\text{syst} (1)_{\alpha_s} (1)_{\fpiPDG}~\MeV ,
\end{equation}
in which the ${\alpha_s}$ uncertainty has become very small.

Using the above results and Eqs.~(\ref{eq:rabbit-a}) and~(\ref{eq:mSI2mMRS}), we obtain the charm and bottom masses in the MRS
scheme:
\begin{align}
    m_{c,\MRS} &= 1392 (6)_\text{stat} (8)_\text{syst} (6)_{\alpha_s} (0)_{\fpiPDG}~\MeV ,
    \label{eq:mc_MRS} \\
    m_{b,\MRS} &= 4749 (14)_\text{stat} (2)_\text{syst} (11)_{\alpha_s} (1)_{\fpiPDG}~\MeV ,
    \label{eq:mb_MRS}
\end{align}
or, adding all errors in quadrature, $m_{c,\MRS}=1392(12)$~MeV and $m_{b,\MRS}=4749(18)$~MeV.
Similar to the stability shown in Fig.~\ref{fig:stability_loop}, the ratio $m_\MRS/\mbar$ is very stable.
For $\alpha_s=0.22$ and three flavors of massless quarks,
\begin{equation}
    m_\MRS/\mbar = (1.133, 1.131, 1.132, 1.132) 
    \label{eq:MRS-MSbar}
\end{equation}
at one through four loops, while
\begin{equation}
    m_\pole/\mbar = (1.093, 1.143, 1.183, 1.224),
\end{equation}
omitting in both cases the charm sea-quark contribution $\Delta m_{(c)}$ for simplicity.

If we use the PDG's estimate of the uncertainty in $\alpha_\MSbar$ instead of that in Eq.~(\ref{eq:alpha5GeV}), then each
uncertainty associated with $\alpha_s$ increases by about 50\%~or so, namely to 0.78~MeV, 0.039~MeV, and 0.018~MeV, for the
strange-, down-, and up-quark masses; 6.0~MeV and 14~MeV for $m_{c,\MSbar}(3~\GeV)$ and $\mbar_c$; and 12~MeV for $\mbar_b$.

\subsection{HQET matrix elements}
\label{subsec:hqet}

The EFT fit directly yields results for the HQET matrix elements.
With the minimal renormalon subtraction, our result
\begin{equation}
    \Lambdabar_\MRS   = 555  (25)_\text{stat}  (8)_\text{syst} (16)_{\alpha_s}  (1)_{\fpiPDG}~\MeV ,
    \label{eq:Lambadbar_MRS}
\end{equation}
is renormalon-free.
This value corresponds to light valence mass $\half(m_u+m_d)$.
The kinetic and chromomagnetic matrix elements are
\begin{align}
    \mu_\pi^2  	     &= 0.05 (16)_\text{stat} (13)_\text{syst} (06)_{\alpha_s} (00)_{\fpiPDG}~\GeV^2 ,
    \label{eq:mu_pi_sq} \\
    \mu_G^2(m_b)     &= 0.38 (01)_\text{stat} (01)_\text{syst} (00)_{\alpha_s} (00)_{\fpiPDG}~\GeV^2 . 
    \label{eq:mu_G_sq}
\end{align}
This value for $\mu_G^2(m_b)$ cannot be considered as a pure lattice-QCD determination because, as discussed in
Sec.~\ref{sec:SChPT}, the prior for $\mu_G^2(m_b)\sim0.35(7)~\GeV^2$ comes from the $B$-meson hyperfine splitting.
The definition of $\mu_\pi^2$ used here still has a renormalon ambiguity of order $\LamQCD^2$,
although it is expected to be small~\cite{Neubert:1996zy,Brambilla:2017mrs}.
In any case, the result in Eq.~(\ref{eq:mu_pi_sq}) cannot be directly compared with results in the ``kinetic''
scheme~\cite{Uraltsev:1996rd,Uraltsev:2001ih}, where $\mu_\pi^2\approx\mu_G^2$ is expected~\cite{Uraltsev:2003ye} and
roughly holds~\cite{Alberti:2014yda,Gambino:2017vkx}.
We checked whether our $\chi^2$ function could be consistent with such an outcome by starting the fit at $\mu_\pi^2=0.35~\MeV^2$,
but found the same minimum as in Eq.~(\ref{eq:mu_pi_sq}).
We also have tried changing the prior for $\mu_\pi^2$ from $(0 \pm 0.36)~\GeV^2$ to $(0.35 \pm 0.36)~\GeV^2$, in which case $\chi^2$
is minimized for $\mu_\pi^2=0.09(16)~\GeV^2$ and $\mu_G^2(m_b)=0.39(1)~\GeV^2$, where the errors are statistical only here.

To compare Eq.~(\ref{eq:Lambadbar_MRS}) with the RS scheme at a given factorization scale $\nu_f$,
one can use~\cite{Brambilla:2017mrs}
\begin{equation}
    \Lambdabar_\RS(\nu_f) = \Lambdabar_\MRS + \cJ_\MRS(\nu_f) ,
    \label{eq:Lambadbar:MRS2RS}
\end{equation}
with the function $\cJ_\MRS$ given in Eq.~(2.37) of Ref.~\cite{Brambilla:2017mrs}.
Setting $\nu_f=1$~GeV, we find
\begin{equation}
    \Lambdabar_\RS(1~\GeV) = 639 (25)_\text{stat}  (8)_\text{syst} (24)_{\alpha_s} (1)_{\fpiPDG}~\MeV .\label{eq:Lambadbar_RS}
\end{equation}
The uncertainty associated with $\alpha_s$ is larger here than for $\Lambdabar_\MRS$, because $\cJ_\MRS(\nu_f)$ in
\eq{Lambadbar:MRS2RS} depends on $\alpha_s(\nu_f)$.
Our result for $\Lambdabar_\RS(1~\GeV)$ agrees with $\Lambdabar_\RS(1~\GeV)=659$~MeV (no error quoted)~\cite{Pineda:2001zq} and
623~MeV (after rough conversion of a result in the ``RS$'$'' scheme)~\cite{Bali:2003jq}, which are obtained from the
$B$-meson mass and the RS mass for the bottom quark.

For future phenomenological studies, Table~\ref{tab:correlation_matrix_ME} in the Appendix provides
the correlation matrix of the MRS masses of the charm and bottom quarks with the HQET matrix elements
$\Lambdabar_\MRS$, $\mu_\pi^2$ and $\mu_G^2(m_b)$.

\subsection{Flavor splittings}
\label{subsec:flavor-splittings}

We use the $D_s$- and $B_s$-meson masses as experimental input to set the $c$- and $b$-quark masses.
Comparing the output of the fit at $m_x=m_d$ with $m_x=m_s$, we obtain the flavor splittings
\begin{align}
    M_{D_s}-M_{D^+} &= 97.9 (0.2)_\text{stat} (0.2)_\text{syst} (0.0)_{\alpha_s} (0.1)_{\fpiPDG} (0.5)_{g_\pi}~\MeV ,
    \label{eq:diff_Ds_Dp} \\
    M_{B_s}-M_{B^0} &= 87.1 (0.4)_\text{stat} (1.0)_\text{syst} (0.0)_{\alpha_s} (0.1)_{\fpiPDG} (0.5)_{g_\pi}~\MeV .
    \label{eq:diff_Bs_B0}
\end{align}
These results agree with the experimental values~\cite{Olive:2016xmw}
\begin{align}
    (M_{D_s}-M_{D^+})^\text{expt} &= 98.69(5)~\MeV ,
    \label{eq:diff_Ds_Dp-PDG} \\
    (M_{B_s}-M_{B^0})^\text{expt} &= 87.3 (2)~\MeV .
    \label{eq:diff_Bs_B0-PDG}
\end{align}
In these combinations of meson masses, the leading-order electromagnetic contributions cancel.
The last uncertainty here stems from the significant changes found in the alternate fits
with $g_\pi$ fixed to 0.45 or to~0 (the polynomial fit).

In a similar vein, we can set the quark masses to $m_\vt=m'_l=m'_s=0$ to obtain the SU(3) chiral limit of charmed and 
$b$-flavored mesons, or set $m_\vt=m'_l=0$ and leave $m'_s=m_s$ to obtain the SU(2) chiral limit.
The results are
\begin{align}
    M_D^\text{SU(3)} &= 1842.7 (2.2)_\text{stat} (1.4)_\text{syst} (0.1)_{\alpha_s} (0.1)_{\fpiPDG} (1.6)_{g_\pi}~\MeV \\
    M_D^\text{SU(2)} &= 1862.3 (0.3)_\text{stat} (1.3)_\text{syst} (0.0)_{\alpha_s} (0.1)_{\fpiPDG} (0.1)_{g_\pi}~\MeV
\end{align}
for the $D$ system, and
\begin{align}
    M_B^{\text{SU(3)}} &= 5245.1 (3.2)_\text{stat} (2.7)_\text{syst} (0.1)_{\alpha_s} (0.1)_{\fpiPDG} (2.1)_{g_\pi}~\MeV \\
    M_B^{\text{SU(2)}} &= 5272.9 (0.5)_\text{stat} (1.3)_\text{syst} (0.0)_{\alpha_s} (0.1)_{\fpiPDG} (0.1)_{g_\pi}~\MeV
\end{align}
for the $B$ system.
This information can be combined with Table~XII of Ref.~\cite{Bazavov:2017fBD}, to derive decay constants from the values of 
$\Phi=\sqrt{M}f$ tabulated there.

\subsection{Low-energy constants in \boldmath\texorpdfstring{\hmchpt}{HMChPT}}
\label{subsec:LECs}
Reference~\cite{Bernard:2017npd} uses the LECs $\lambda_1$ and $\lambda'_1$ obtained in this work.
In particular, the values used are those for $D$ mesons, which come from the simple, polynomial
analysis without chiral expressions, \ie $g_\pi=0$:
\begin{align}
    \breve{\lambda}_{1,D}  &= 0.218 (2)~\GeV^{-1} ,\label{eq:lambda_1_Taylor:D}\\
    \breve{\lambda}'_{1,D} &= 0.037(13)~\GeV^{-1} ,\label{eq:lambda_1_prime_Taylor:D}
\end{align}
where the errors are statistical only, which suffices for \rcite{Bernard:2017npd}.
Here, the breve is a reminder that finite-mass corrections to the LECs in the \hmchpt\ Lagrangian are included.
From the experimental data for the flavor splittings of $D$ mesons, one finds
$\breve\lambda_1\approx0.2~\GeV^{-1}$~\cite{Bazavov:2011aa}.

\section{Summary, Comparisons, and Outlook}
\label{sec:Outlook}

The results presented in Sec.~\ref{sec:Results} show that the new HQET-based method, developed here and in
Ref.~\cite{Brambilla:2017mrs}, is both qualitatively and quantitatively successful.
The qualitative success relies on the clean separation of scales provided by HQET with the MRS definition of the heavy-quark mass,
while the quantitative success relies on the high statistics of the MILC Collaboration's HISQ
ensembles~\cite{Bazavov:2010ru,\rHISQrCONFIGS,milc_hisq}, all 24 of which have been employed here.
Also relevant to the success of the method is the availability of the order-$\alpha_s^5$ perturbation theory for the running of the
quark mass~\cite{Baikov:2014qja} and strong coupling~\cite{Baikov:2016tgj}, and the order-$\alpha_s^4$ coefficient linking the
\MSbar\ mass to the pole mass and, hence, the MRS mass~\cite{Marquard:2015qpa,Marquard:2016dcn}.
These features are not (yet) shared by other determinations of quark masses using lattice QCD.
Although the HQET method separates the heavy-quark scale from the QCD scale, mass ratios determined in the course of this work and
Ref.~\cite{Bazavov:2017fBD} yield results for all quarks except the top quark.

\begin{figure}[b]
    \includegraphics[trim={6.5em 0em 1em 1em},clip,width=0.495\textwidth]{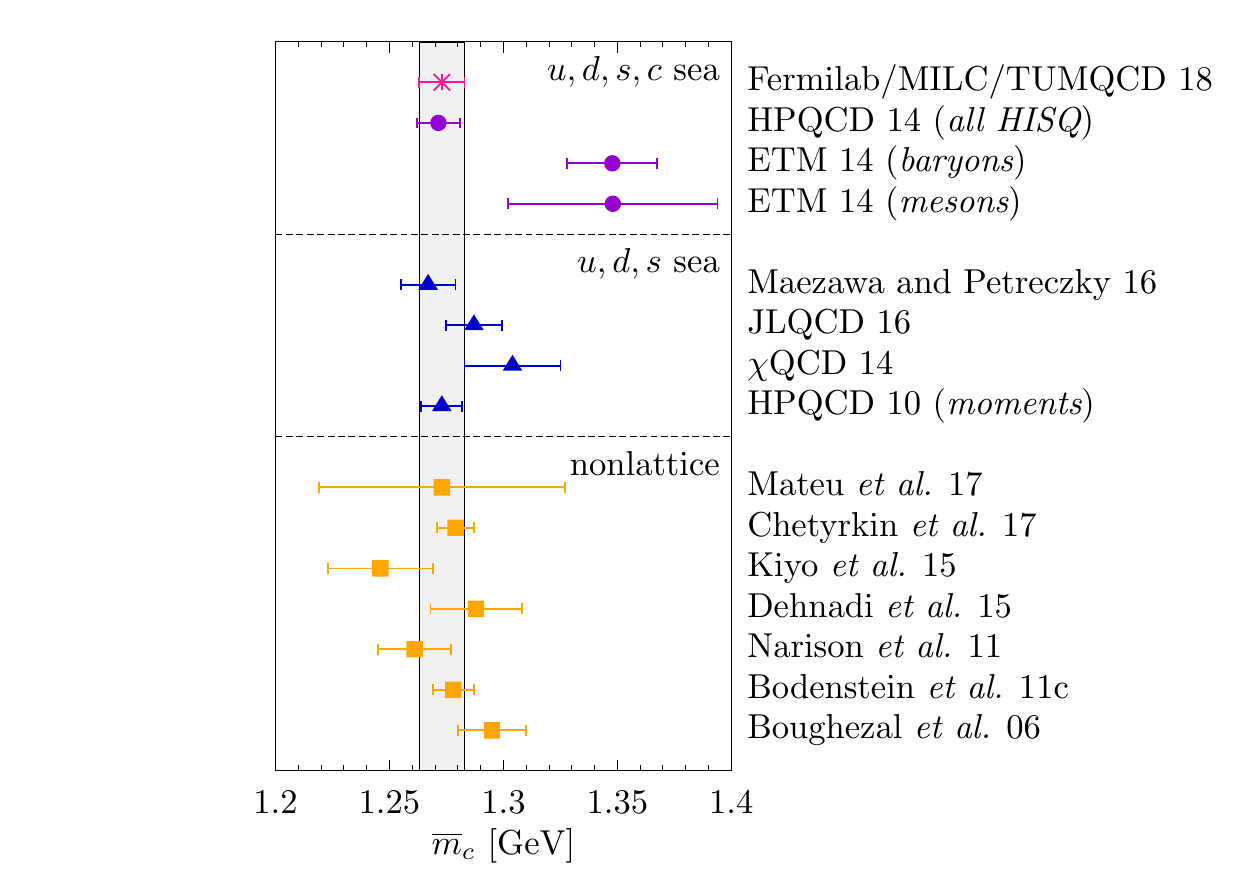} \hfill
    \includegraphics[trim={6.5em 0em 1em 1em},clip,width=0.495\textwidth]{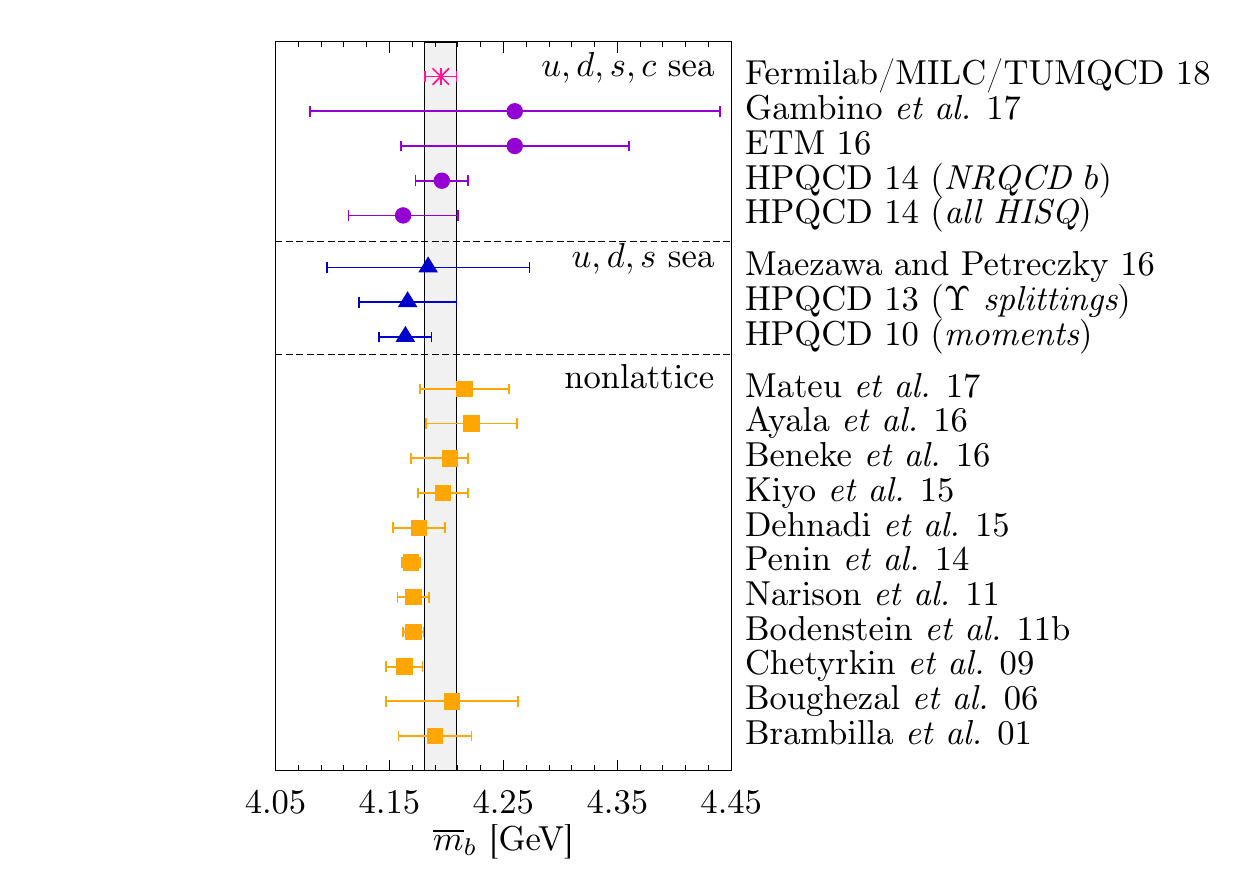}
    \caption[fig:mcmb]{Comparison of $\mbar_c$ (left) and $\mbar_b$ (right) to other results from lattice QCD and from nonlattice
    methods.
    Our result is shown as a magenta burst, with the gray band showing how it compares directly with the other results.
    The labels refer to \linebreak
        Fermilab/MILC/TUMQCD 18 (this work);  
        HPQCD~14 (\emph{all HISQ})~\cite{Chakraborty:2014aca};
        ETM~14 (\emph{baryons}) \cite{Alexandrou:2014sha};
        ETM~14~(\emph{mesons}) \cite{Carrasco:2014cwa}; 
        Maezawa and Petreczky~16 \cite{Maezawa:2016vgv};
        JLQCD~16 \cite{Nakayama:2016atf};
        $\chi$QCD~14 \cite{Yang:2014sea};
        HPQCD~10 (\emph{moments}) \cite{McNeile:2010ji};
        Mateu      \emph{et al.}~17 \cite{Mateu:2017hlz};
        Chetyrkin  \emph{et al.}~17 \cite{Chetyrkin:2017lif};
        Kiyo       \emph{et al.}~15 \cite{Kiyo:2015ufa};
        Dehnadi    \emph{et al.}~15 \cite{Dehnadi:2015fra};
        Narison~11 \cite{Narison:2011xe};
        Bodenstein \emph{et al.}~11c \cite{Bodenstein:2011ma};
        Boughezal  \emph{et al.}~06 \cite{Boughezal:2006px};
        Gambino    \emph{et al.}~17 \cite{Gambino:2017vkx};
        ETM~16 \cite{Bussone:2016iua};
        HPQCD~14 (\emph{NRQCD} $b$) \cite{Colquhoun:2014ica};
        HPQCD~13 ($\Upsilon$~\emph{splittings}) \cite{Lee:2013mla};
        HPQCD~10 (\emph{moments}) \cite{McNeile:2010ji};
        Ayala      \emph{et al.}~16 \cite{Ayala:2016sdn};
        Beneke     \emph{et al.}~16 \cite{Beneke:2016oox,*Beneke:2014pta};
        Penin      \emph{et al.}~14 \cite{Penin:2014zaa};
        Bodenstein \emph{et al.}~11b \cite{Bodenstein:2011fv};
        Chetyrkin  \emph{et al.}~09 \cite{Chetyrkin:2009fv};
        and Brambilla  \emph{et al.}~01 \cite{Brambilla:2001qk}.
    }
    \label{fig:mcmb}
\end{figure}

Our results for heavy-quark masses $\mbar_c$ and $\mbar_b$ are compared with other results in the literature in
Fig.~\ref{fig:mcmb}.
Both panels show the most recent lattice-QCD calculation with a complete error budget from each combination of method and
collaboration.
For nonlattice calculations, we also show the most recent result from each method and-or collaboration, but include only those with
perturbative-QCD accuracy of order-$\alpha_s^3$ matching and, if needed, order-$\alpha_s^4$ running.
As noted in Sec.~\ref{sec:Results}, the parametric uncertainty in $\alpha_s$ is one of our largest uncertainties, but, thanks to the
MRS mass, higher-order perturbative corrections are likely to be negligible compared with this and our statistical uncertainty; cf.\
Fig.~\ref{fig:stability_loop} and Eq.~(\ref{eq:MRS-MSbar}).

For $\mbar_c$, the overall agreement is very good, and our result's uncertainty is about the same as those from charmonium
correlators and (continuum) perturbative QCD, using either lattice~\cite{Chakraborty:2014aca} or
experimental~\cite{Chetyrkin:2017lif,Bodenstein:2011ma} data as input.
(References~\cite{Chetyrkin:2017lif} and~\cite{Bodenstein:2011ma} differ in the moments used.)
The difference between our result for $m_{c,\MSbar}(3~\GeV)$ and the recent update from Chetyrkin \emph{et
al.}~\cite{Chetyrkin:2017lif} is~$0.9\sigma$. 
For $\mbar_b$, the overall agreement is good.
The difference between our result and those of Narison~\cite{Narison:2011xe}, Bodenstein \emph{et al.}~\cite{Bodenstein:2011fv}, 
Chetyrkin \emph{et al.}~\cite{Chetyrkin:2009fv}, and Penin and Zerf~\cite{Penin:2014zaa} is 
$1.3\sigma$, 
$1.6\sigma$, 
$1.6\sigma$, 
and $1.7\sigma$, 
respectively.
Such discrepancies among 19 independent results, especially given the importance of systematic uncertainties in all determinations,
should not be seen as alarming.

It is noteworthy that for $\mbar_c=m_{c,\MSbar}(m_{c,\MSbar})$ the result of Ref.~\cite{Chetyrkin:2017lif} is more precise than 
ours, while for $m_{c,\MSbar}(3~\GeV)$ ours is more precise.
In both cases, the error bar runs as dictated by the quark-mass anomalous dimension and beta function.
In addition, the order-$\alpha_s$ coefficient is proportional to $[\ln(3~\GeV/\mu) + c]$.
For the relation between $\mbar$ and $m_\MRS$, $c>0$, so the first-order $\alpha_s$ error vanishes for some $\mu>3~\GeV$.
On the other hand, for the relation between $\mbar$ and moments of the charmonium correlator, $c<0$, so the first-order $\alpha_s$ 
error vanishes for some $\mu<3~\GeV$.
We therefore also provide light, strange, and charm masses at 3~GeV:
\begin{align}
  m_{l,\MSbar}(3~\GeV)  &= 3.072  (13)_\text{stat}  (04)_\text{syst}  (10)_{\alpha_s}  (04)_{\fpiPDG}~\MeV , \\
  m_{u,\MSbar}(3~\GeV)  &= 1.923  (16)_\text{stat}  (32)_\text{syst}  (06)_{\alpha_s}  (02)_{\fpiPDG}~\MeV , \\
  m_{d,\MSbar}(3~\GeV)  &= 4.221  (27)_\text{stat}  (35)_\text{syst}  (14)_{\alpha_s}  (05)_{\fpiPDG}~\MeV , \\
  m_{s,\MSbar}(3~\GeV)  &= 83.49  (36)_\text{stat}  (16)_\text{syst}  (28)_{\alpha_s}  (10)_{\fpiPDG}~\MeV , \\
  m_{c,\MSbar}(3~\GeV)  &= 983.7  (4.3)_\text{stat} (1.4)_\text{syst} (3.3)_{\alpha_s} (0.5)_{\fpiPDG}~\MeV .
\end{align}
In contexts beyond the Standard Model, one needs the masses---that is the Yukawa coupling to the Higgs field---at scales of 100~GeV 
or higher.
Table~\ref{tab:correlation_matrix_masses} in the Appendix provides the correlation matrix
for our charm-quark mass at 3~GeV and quark-mass ratios.

\begin{figure}
    \includegraphics[trim={6em 0em 0em 1em},clip,width=0.495\textwidth]{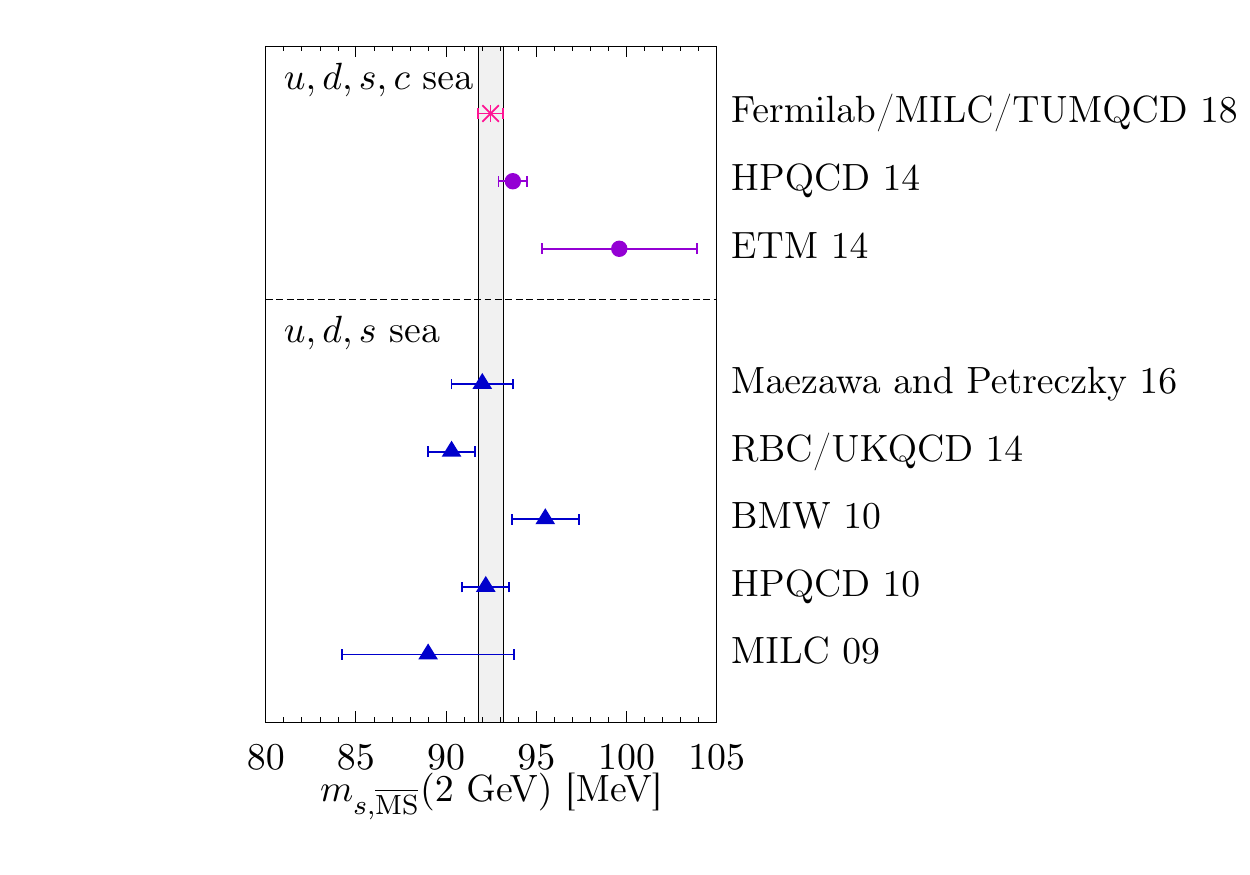} \hfill
    \includegraphics[trim={6em 0em 0em 1em},clip,width=0.495\textwidth]{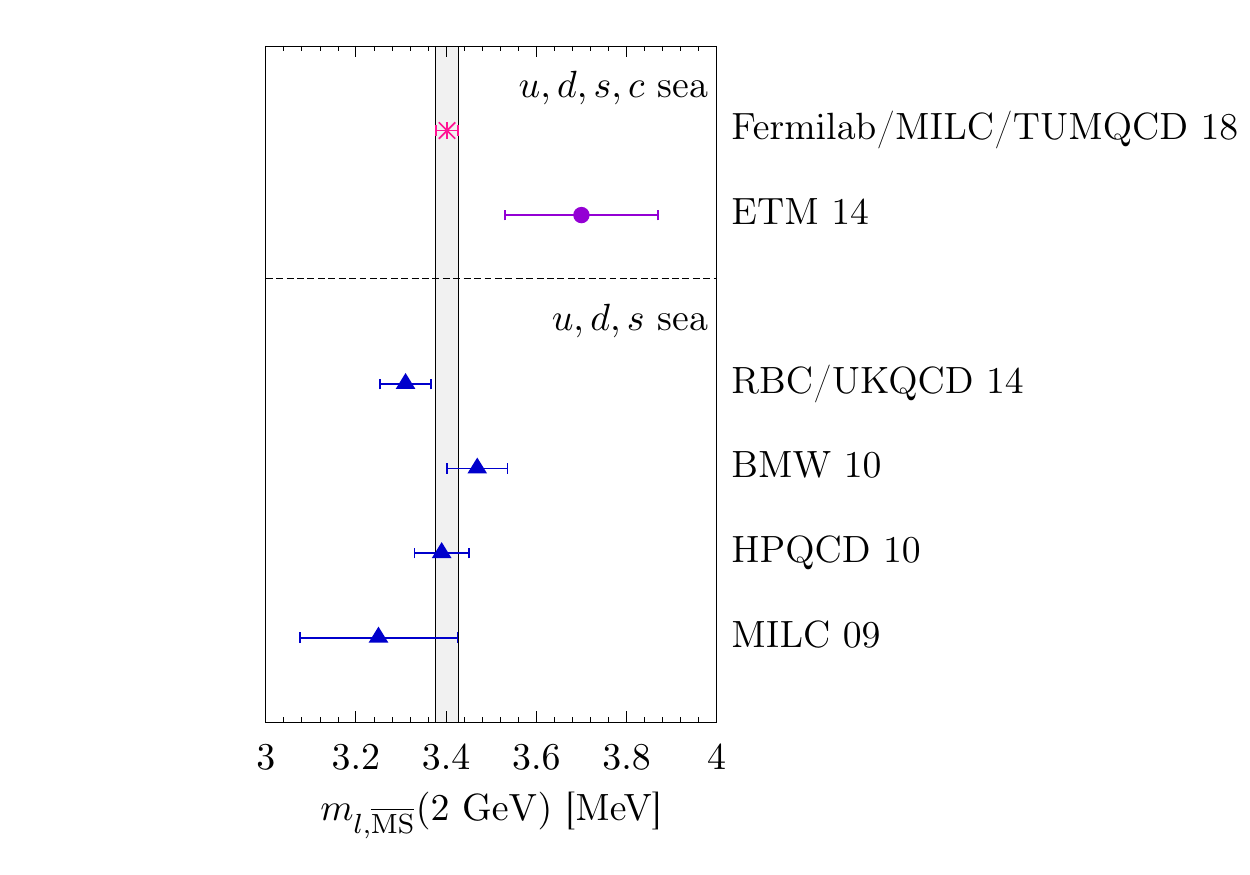}
    \caption{Comparison of $m_{s,\MSbar}(2~\GeV)$ (left) and $m_{ud,\MSbar}(2~\GeV)$ (right) to other results from lattice QCD.
    Our result is shown as a magenta burst, with the gray band showing how it compares directly with the other results.
    The labels refer to
        Fermilab/MILC/TUMQCD 18 (this work);
        HPQCD~14~\cite{Chakraborty:2014aca};
        ETM~14~\cite{Carrasco:2014cwa};
        Maezawa and Petreczky~16 \cite{Maezawa:2016vgv};
        RBC/UKQCD~14~\cite{Blum:2014tka};
        BMW~10~\cite{Durr:2010vn};
        HPQCD~10~\cite{McNeile:2010ji};
        and MILC~09~\cite{Bazavov:2009fk,*Bazavov:2009bb}.
    }
    \label{fig:msmud}
\end{figure}

Our results for light-quark masses are compared with other results from lattice QCD in Fig.~\ref{fig:msmud}.
As above, both panels show the most recent lattice-QCD calculations with a complete error budget from each combination of method and
collaboration.
As can be seen from the plots, and similar comparisons of $m_u$ and $m_d$, ours are the most precise results to date.
Here the precision stems from very precise quark-mass ratios from the pseudoscalar meson spectrum, together with the overall scale
of quark masses from the EFT~fit.
Consequently, the results inherit an uncertainty due to $\alpha_s$, which is largest except in the cases of $m_{d,\MSbar}(2~\GeV)$
and $m_{u,\MSbar}(2~\GeV)$, which have larger statistical and electromagnetic systematic uncertainties from~$m_u/m_d$.

As compelling as these results are, they could be improved in several ways.
First, because the EFT fit controls systematics, the statistical error (after propagation through the fit) is often the
second-largest source of uncertainty, so, as usual, having more data would reduce the error.
The additional data need not be more precise per se: the right panel of Fig.~\ref{fig:mass-vs-mMRSinv} suggests that finer lattice
spacings will be needed.
Second, because the other dominant uncertainty is the parametric error of $\alpha_s$, it would be interesting to carry out a
simultaneous determination of $\alpha_s$ and the quark masses, for example in a combined analysis of heavy-light meson masses and
quarkonium correlators.
Such an analysis would output $\mbar_c$, $\mbar_b$, and $\alpha_s$ \emph{with their correlations}, which would be very convenient
for determining Higgs-boson branching ratio in the Standard Model and extensions thereof.
Third, QCD+QED simulations would eliminate the scheme dependence arising from the matching of QCD+QED to pure QCD.
Finally, the ideal determination of the matrix elements $\mu_\pi^2$ and $\mu_G^2$, and analogous quantities that enter at
order~$1/m_Q^2$ and higher, would require computing heavy-light vector mesons on the lattice,
in addition to the pseudoscalar mesons studied here.
In particular, this would make possible a pure lattice result for $\mu_G^2$, without making use of the experimental information on
the $B$-meson hyperfine splitting.

\acknowledgments

J.K. thanks Jaume Tarr\'us Castell\`a for a useful discussion on nonanalytic terms in \chpt.

Computations for this work were carried out with resources provided by
the USQCD Collaboration,
the National Energy Research Scientific Computing Center,
the Argonne Leadership Computing Facility, 
the Blue Waters sustained-petascale computing project,
the National Institute for Computational Science,
the National Center for Atmospheric Research,
the Texas Advanced Computing Center,
and Big Red II+ at Indiana University.
USQCD resources are acquired and operated thanks to funding from the Office of Science of the U.S. Department of Energy.
%
The National Energy Research Scientific Computing Center is a DOE Office of Science User Facility supported by the
Office of Science of the U.S. Department of Energy under Contract No.\ DE-AC02-05CH11231.
%
An award of computer time was provided by the Innovative and Novel Computational Impact on Theory and Experiment (INCITE) program.
This research used resources of the Argonne Leadership Computing Facility, which is a DOE Office of Science User Facility supported
under Contract DE-AC02-06CH11357.
%
The Blue Waters sustained-petascale computing project is supported by the National Science Foundation (awards OCI-0725070 and
ACI-1238993) and the State of Illinois.
Blue Waters is a joint effort of the University of Illinois at Urbana-Champaign and its National Center for Supercomputing
Applications.
This work is also part of the ``Lattice QCD on Blue Waters'' and ``High Energy Physics on Blue Waters'' PRAC allocations supported
by the National Science Foundation (award numbers 0832315 and 1615006).
This work used the Extreme Science and Engineering Discovery Environment (XSEDE), which is supported by National Science Foundation
grant number ACI-1548562~\cite{XSEDE_REF}.
Allocations under the Teragrid and XSEDE programs included resources at the National Institute for Computational Sciences (NICS) at
the Oak Ridge National Laboratory Computer Center, the Texas Advanced Computing Center and the National Center for Atmospheric
Research, all under NSF teragrid allocation TG-MCA93S002.
Computer time at the National Center for Atmospheric Research 
was provided by NSF MRI Grant CNS-0421498, NSF MRI Grant CNS-0420873, NSF MRI Grant CNS-0420985, NSF sponsorship of the National
Center for Atmospheric Research, the University of Colorado, and a grant from the IBM Shared University Research (SUR) program.
Computing at Indiana University is supported by Lilly Endowment, Inc., through its support for the Indiana University Pervasive
Technology Institute.

This work was supported in part by the U.S.\ Department of Energy under grants
No.~DE-FG02-91ER40628 (C.B., N.B.),
No.~DE-FC02-12ER41879 (C.D.),
No.~DE{-}SC0010120 (S.G.),     
No.~DE-FG02-91ER40661 (S.G.),
No.~DE-FG02-13ER42001 (A.X.K.),
No.~DE{-}SC0015655 (A.X.K.), 
No.~DE{-}SC0010005 (E.T.N.),
No.~DE-FG02-13ER41976 (D.T.);
by the U.S.\ National Science Foundation under grants
PHY14-14614 and PHY17-19626 (C.D.),
PHY14-17805~(J.L.),
and PHY13-16748 and PHY16-20625 (R.S.);
by the MINECO (Spain) under grants FPA2013-47836-C-1-P and FPA2016-78220-C3-3-P (E.G.);
by the Junta de Andaluc\'{\i}a (Spain) under grant No.\ FQM-101 (E.G.);
by the DFG cluster of excellence ``Origin and Structure of the Universe'' (N.B., A.V.);
by the UK Science and Technology Facilities Council (J.K.);
by the German Excellence Initiative and the European Union Seventh Framework Program under grant agreement No.~291763 as well as 
the European Union's Marie Curie COFUND program (J.K., A.S.K.).
Brookhaven National Laboratory is supported by the United States Department of Energy, Office of Science, Office of High Energy
Physics, under Contract No.\ DE{-}SC0012704.
This document was prepared by the Fermilab Lattice, MILC, and TUMQCD Collaborations using the resources of the Fermi National
Accelerator Laboratory (Fermilab), a U.S.\ Department of Energy, Office of Science, HEP User Facility.
Fermilab is managed by Fermi Research Alliance, LLC (FRA), acting under Contract No.\ DE-AC02-07CH11359.

\appendix
\section{Correlation matrices}
\label{app:appendix}

We report in Table~\ref{tab:correlation_matrix_ME} the correlation matrix
of the MRS masses of the charm and bottom quarks with the HQET matrix elements,
and in Table~\ref{tab:correlation_matrix_masses} the correlation matrix for our charm-quark mass and quark-mass ratios.
Knowledge of these correlations may be useful for future phenomenological studies.

\begin{table}
\newcommand{\h}{\phantom{x}}
\caption{Correlation matrix between the MRS masses of the charm and bottom quarks and HQET matrix elements;
entries are symmetric across the diagonal.
The last row gives the central value and total uncertainty (added in quadrature) of each quantity.}
\label{tab:correlation_matrix_ME}
\begin{tabular}{l|ccccc}
\hline\hline
		    & \h$m_{c,\MRS}$\h 	& \h$m_{b,\MRS}$\h &\h$\Lambdabar_\MRS$\h& $\mu_\pi^2$    & $\mu_G^2(m_b)$ \\
\hline
$m_{c,\MRS}$\h	    & \h1\h 	     & 		      &	 		&		 & \\
$m_{b,\MRS}$\h	    & \h0.72437434\h & \h1\h 	      & 		&		 & \\
$\Lambdabar_\MRS$\h & \h0.14207020\h &$-$0.26823406\h & \h1\h 		&		 & \\
$\mu_\pi^2$\h	    &$-$0.01634290\h & \h0.64044459\h & \h-0.60154065\h & \h1\h 	 & \\
$\mu_G^2(m_b)$\h    &$-$0.28580359\h & \h0.10674678\h & \h-0.12545531\h & \h0.57546979\h & \h1 \\
\hline
    & \h1392(11)~MeV\h & \h4749(18)~MeV\h & \h555(31)~MeV\h & \h0.05(22)~GeV$^2$\h & \h0.38(2)~GeV$^2$ \\
\hline\hline
\end{tabular}
\end{table}

\begin{table}
\newcommand{\h}{\phantom{x}}
\caption{Correlation matrix between $m_{c,\MSbar}(3~\GeV)$ and quark mass ratios;
entries are symmetric across the diagonal.
The last row gives the central value and total uncertainty (added in quadrature) of each quantity.} 
\label{tab:correlation_matrix_masses}
\begin{tabular}{l|ccccc}
\hline\hline
	    & \h$m_{c,\MSbar}(3~\GeV)$\h   & $m_b/m_c$      & $m_s/m_c$      & $m_d/m_c$       & $m_u/m_c$ \\
\hline
$m_{c,\MSbar}(3~\GeV)$\h & \h1\h  	   &  		    & 		     & 		       & \\
$m_b/m_c$\h 		 & $-$0.58607809\h & \h1\h 	    & 		     & 		       & \\
$m_s/m_c$\h 		 & $-$0.11425384\h & \h0.45502225\h & \h1\h 	     & 		       & \\
$m_d/m_c$\h 		 &  \h0.14213251\h & \h0.04855992\h & \h0.43609054\h & \h1\h 	       & \\
$m_u/m_c$\h 		 & $-$0.16516954\h & \h0.23627864\h & \h0.47252309\h & $-$0.32724921\h & \h1 \\
\hline
	& \h983.7(5.6)~MeV\h & \h4.578(8)\h & \h0.08487(18)\h & \h0.004291(39)\h & \h0.001955(37) \\
\hline\hline
\end{tabular}
\end{table}

\bibliographystyle{apsrev4-1}
\bibliography{References}

\end{document}